\documentclass[
	fontsize=12pt,          
	headsepline=false,       
	footsepline=false, 		
	titlepage=true,
    DIV=12                	
]{scrartcl}

\usepackage[section]{placeins}
\usepackage{textcomp} 
\usepackage[T1]{fontenc}                 
\usepackage[utf8]{inputenc}            
\usepackage[USenglish, american ]{babel}  
\usepackage{graphicx,wrapfig,lipsum}
\usepackage[paper=a4paper,left=29mm,right=29mm,top=30mm,bottom=30mm]{geometry}
\usepackage{tikz}
\usetikzlibrary{shapes.geometric, arrows}
\usetikzlibrary{decorations.pathreplacing}
\usepackage{prodint}
\usepackage{pdfpages}
\usepackage[nohyperlinks]{acronym}

\usepackage{graphicx}
\usepackage{mwe}
\usepackage{subfig}
\usepackage[authoryear]{natbib}
\usepackage{float}
\usepackage{csquotes}
\usepackage[all]{nowidow}
\usepackage{xparse}
\usepackage{url}
\usepackage{diagbox}
\usepackage{amsmath, amsfonts,mathtools, amssymb, bbm, bm,nicefrac}
\usepackage{xcolor}
\usepackage{authblk}
\usepackage[onehalfspacing]{setspace} 
\usepackage{csquotes}
\usepackage{scrlayer-scrpage}
\ihead{}				
\ohead{\headmark}					
\ifoot{}							
 	\ofoot{}							

\usetikzlibrary{shapes,arrows,shadows, positioning, calc}
\usepackage[all]{nowidow}
\usepackage{booktabs} 	
\usepackage{longtable} 
\usepackage[hidelinks]{hyperref} 
\usepackage{enumitem}
\allowdisplaybreaks

\renewcommand{\title}[1]{ \noindent{\centering \Large \textbf{ #1 } \\} }
\newcommand{\inst}[1]{\textsuperscript{#1}}
\newcommand{\institute}[1]{{\centering \footnotesize{#1}} \vspace{2ex}}

\begin{document}

\title{Comparison of Estimators\\
for Multi-State Models\\ in Potentially Non-Markov Processes}

\begin{center}
		Carolin Drenda\inst{1}, Dennis Dobler\inst{2,*}\let\thefootnote\relax\footnote{* Authors listed in alphabetic order.}, Merle Munko\inst{3,*$\dagger$}\let\thefootnote\relax\footnote{$\dagger$ Corresponding author. Email address: \url{merle.munko@ovgu.de}}, Andrew Titman\inst{4,*}
\end{center}

	\institute{
        \inst{1} TU Dortmund University; Dortmund (Germany)\newline
        \inst{2} RWTH Aachen University; Aachen (Germany)\newline
		\inst{3} Otto-von-Guericke University Magdeburg; Magdeburg (Germany)\newline
        \inst{4} Lancaster University; Lancaster (United Kingdom)
	}

\hrule
\vspace{1cm}
\noindent\textbf{Abstract}
Various estimators for modelling the transition probabilities in multi-state models have been proposed, e.g.,
the Aalen-Johansen estimator, the landmark Aalen-Johansen estimator, and a hybrid Aalen-Johansen estimator.
While the Aalen-Johansen estimator is generally only consistent under the rather restrictive Markov assumption, the landmark Aalen-Johansen estimator can handle non-Markov multi-state models.
However, the landmark Aalen-Johan\-sen estimator leads to a strict data reduction and, thus, to an increased variance.
The hybrid Aalen-Johansen estimator serves as a compromise by, firstly, checking with a log-rank-based test whether the Markov assumption is satisfied. Secondly, landmarking is only applied if the Markov assumption is rejected.
In this work, we propose a new hybrid Aalen-Johansen estimator which uses a Cox model instead of the log-rank-based test to check the Markov assumption in the first step.
Furthermore, we compare the four estimators in an extensive simulation study across Markov, semi-Markov, and distinct non-Markov settings.
In order to get deep insights into the performance of the estimators, we consider four different measures: bias, variance, root mean squared error, and coverage rate.
Additionally, further influential factors on the estimators such as the form and degree of non-Markov behaviour, the different transitions, and the starting time are analysed.
The main result of the simulation study is that the hybrid Aalen-Johansen estimators yield favourable results across various measures and settings.

\noindent\textbf{Keywords:} 
Aalen-Johansen estimator, frailty models, hybrid Aalen-Johansen estimator, non-Markov model, semi-Markov model, simulation study.

\section{Introduction}

Multi-state models extend classical survival models by involving longitudinal data with multiple transitions between different states over time. States represent different situations or conditions of the subjects, like being disease-free or in different stages of a disease. Rather than solely focusing on the time at which the disease progression happens, researchers may also be interested in understanding how and when patients traverse through different stages of the disease. In these cases standard survival analysis might be too restrictive. To address the complexity of such scenarios, multi-state models have emerged as powerful tools in event history analysis. 
Some examples are \citet{multi_bei}, examining the effectiveness of treatment sequences for follicular lymphoma, or \citet{multibsp2}, exploring prognostic factors in the progression of breast cancer. 

Unlike traditional survival analysis methods, multi-state models enable the modelling and analysis of multiple events simultaneously. These models accommodate any finite number of states with events serving as transitions between them. The specific structure of the model depends on the characteristics of the data under analysis and the research question being investigated. Among these models, Markov models stand out, since they are notably simpler in terms of probability compared to other models. In this type of model, transitions to future states depend solely on the current state and not the past.
One key aspect in the analysis of multi-state models is the estimation of the transition probabilities, as they allow for long-term predictions \citep{hougaard1999multi}.
A consistent estimator for the probability in the Markov case is the Aalen-Johansen estimator introduced by \citet{aalen1978empirical}. 
The Markov assumption, however, might be unrealistic for real-world data, as it imposes strict conditions. For instance, when examining long term work and sick leave, 
 it is reasonable to assume that individuals with longer periods of working in their jobs may be in more stable positions. Therefore, they potentially prolong their employment duration.
 In cases where the Markov property is not met, such as non-Markov scenarios, the results obtained from the Aalen-Johansen estimator may be biased. This raises a concern, as the Markov property is often not verified \citep{maltzahn2021hybrid}. 

To address the challenges of non-Markov scenarios, several estimators have been proposed. One of them is the landmark Aalen-Johansen estimator by \citet{spitoni2018non}. This approach has proven effectiveness in handling non-Markov behaviour across diverse settings and has shown superiority over alternative methods. It achieves this by employing landmarking, where the data is stratified and only select subjects are used for each transition. While this feature enables consistency in non-Markov cases, it also has a drawback: the strict reduction of data, which can lead to an unreliable point estimate. This limitation becomes particularly pronounced when not a lot of data is available, such as when the overall sample size is small or when some transitions are less common within a model \citep{maltzahn2021hybrid}. 
 Moreover, in the context of clinical trials, the reduction of each patient also implies an economic loss \citep{rohrig2010sample}.

A solution to that problem is the hybrid Aalen-Johansen estimator proposed by \citet{maltzahn2021hybrid}. Their approach involves examining each transition in the model to determine whether the Markov assumption is violated and, if so, whether that is not negligible. Only in those cases where Markov behaviour cannot be justified, the landmarking is applied, leading to less reduction in data. This approach has been particularly successful in addressing partially non-Markov multi-state models, where some transitions are considered to be Markov and others not. In the cases used in their simulation study, this estimator demonstrated a more precise estimation by incorporating more subjects compared to the landmark Aalen-Johansen estimator with minimal increase in bias.

In this work, an adjustment of the hybrid Aalen-Johansen estimator is proposed.
In the original hybrid estimator, a log-rank-based test is applied to check the Markov assumption for each transition. In this new version, a Cox model is applied instead. 
Given the effectiveness of this method to detect non-Markov behaviour across a wide range of scenarios, as demonstrated by \citet{titman2022general}, the aim is to improve the accuracy of the testing component to achieve even better estimation results.

In a simulation study utilising the illness-death model with recovery, the four introduced estimators are compared across various controlled settings. These settings encompass different aspects of Markov, semi-Markov and non-Markov behaviour. 
One primary objective is to analyse the performance of each estimator in those settings. Further factors, like the starting time or degree of non-Markov behaviour, are varied and their influence on the estimation is examined. Additionally, particular emphasis is placed on evaluating the newly introduced estimator. Hence, the secondary aim is to assess whether the new estimator represents an improvement over the existing hybrid estimator as well as over the other two estimators. 

The remainder is structured as follows. Section \ref{sec:intro} offers an overview of the theory surrounding multi-state models. 
In Section \ref{sec:test}, methods for testing the Markov property are explained. This includes the Cox model and the log-rank-based test by \citet{titman2022general}.
Section \ref{sec:data} discusses the set-up and data generation process of the simulation study. 
In Section \ref{sec:res}, the results from the simulation study are presented, comparing the estimators under different aspects.
Lastly, Section \ref{sec:conl} summarises the main insights from the study. 

\section{Review of Multi-State Models}
\label{sec:intro}





\begin{figure}[tb]
\begin{center}
\tikzstyle{block} = [rectangle, draw, fill=gray!30, 
    text width=8em, text centered, rounded corners, minimum height=3em, drop shadow]
\tikzstyle{arrow} = [thick,->,>=stealth]
\scalebox{0.8}{
\begin{tikzpicture}[node distance=6cm]
    \node [block] (death) {State 1: Alive};
    \node [block, right of=death] (healthy) {State 2: Death};
    \draw [arrow] (death) -- node[above] {$\alpha(t)$} (healthy);
\end{tikzpicture}
}
\end{center}
\caption{Classical survival model.}
\label{fig:survival model}
\end{figure}
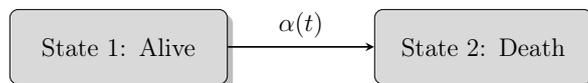

Event history analysis is used to analyse the time until the occurrence of an event. An \textit{event} represents the transition from one state to another.
Classical survival analysis considers one specific event per subject.
Here, the outcome variable is the survival time, which refers to the time progressing from an initiating state (state 1) to a state of interest (state 2). 
A typical example from biostatistics for state 1 is the representation of an individual being alive, while state 2 denotes the individual being deceased. This model can be graphically displayed, as shown in Figure \ref{fig:survival model}, where the boxes illustrate the states and the arrow depicts the transition. 
In more complex cases, where several events per subject are of interest, multi-state models are used to represent the event history data. 


\subsection{Multi-State Model}
\label{sec:multi-state}

In a general multi-state model, there are possibly multiple events of interest, allowing subjects to transition between more than two states. 
To accommodate this, define $\mathcal{S} = \{1,...,k\}$, $k\geq2$, as \textit{state space}. 
It contains a finite number of potential states that a subject can occupy. 
A subject can only be in one state at any given time. 

\begin{figure}[bt]
    \centering
\tikzstyle{block} = [rectangle, draw, fill=gray!30, 
    text width=8em, text centered, rounded corners, minimum height=3em, drop shadow]
\tikzstyle{arrow} = [thick,->,>=stealth]
\scalebox{0.8}{
\begin{tikzpicture}[node distance=5cm]
    \node [block] (healthy) {State 1: Healthy};
    \node [block, right=2cm of death] (illness) {State 2: Illness};
    \node [block, below=1.5cm of $(healthy)!0.5!(illness)$] (death) {State 3: Death};
    \draw [arrow] ([yshift=-0.4cm] healthy.north east) -- node[above] {$\alpha_{12}(t)$} ([yshift=-0.4cm] illness.north west);
    \draw [arrow] ([yshift=0.4cm] illness.south west) -- node[below] {$\alpha_{21}(t)$} ([yshift=0.4cm] healthy.south east);
	\draw [arrow] (healthy.south) -- node[left,xshift=-3mm] {$\alpha_{13}(t)$} (death);
    \draw [arrow] (illness.south) -- node[right, xshift=2mm] {$\alpha_{23}(t)$} (death);
\end{tikzpicture}
}
    \caption{Illness-death model with recovery.}
    \label{fig:illness_death}
\end{figure}
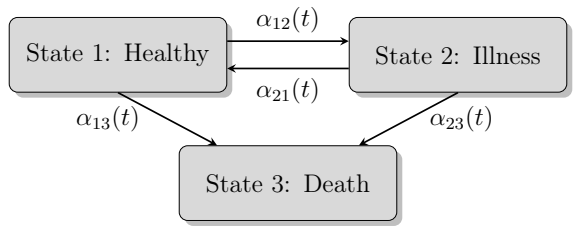

Figure \ref{fig:illness_death} illustrates an example of a multi-state model with three states, known as the illness-death model with recovery. 
The model consists of three states, with the state space denoted as $\mathcal{S} = \{1, 2, 3\}$ and four possible transitions between them. 

To observe a subject's state occupancy over time, a \textit{multi-state process} is defined as a stochastic process \[\bigl(X(t), ~t \in \mathcal{T}\bigr),\] where $X(t) \in \mathcal{S}$ 
and the time interval is $\mathcal{T} = [0, \tau)$, $\tau \leq + \infty$. 
This is a right-continuous piecewise constant process. 
The process generates the natural filtration \[(\mathcal{X}_t)_{t\in\mathcal T} := \left(\sigma (X(u), u \leq t )\right)_{_{t\in\mathcal T}}\]
known as the \textit{history}. 
Then, the \textit{transition probability} is defined as 
\begin{align*}
P_{hj}(s,t\mid \mathcal{X}_{s-}) := \mathbb{P}\Bigl(X(t) = j |X(s) = h, \mathcal{X}_{s-}\Bigr),
\end{align*}  
with $h,~j \in \mathcal{S},~ s,~ t~ \in \mathcal{T}, ~s\leq t$, and $s-$ denotes the left-hand limit of time $s$. 
Consequently, the \textit{transition probability matrix}
\begin{align*}
\mathbf{P}(s,t\mid \mathcal{X}_{s-}) = \Bigl(P_{hj}(s,t\mid \mathcal{X}_{s-})\Bigr)_{h,j\in\mathcal S}  \in [0,1]^{k\times k}, 
\end{align*} aggregates the probabilities of transitioning from one to another state for all state combinations. 
The multi-state versions of the hazard rate function are the \textit{transition intensities} \begin{align*}
\alpha_{hj}(t\mid \mathcal{X}_{s-}) := \lim_{d t \searrow 0} \frac{P_{hj}(t, t + d t\mid \mathcal{X}_{s-})}{d t}.
\end{align*}
It describes the instantaneous risk of transition from state $h$ to $j$. 
All transition intensities can be summarised by the matrix function $\bm{\alpha}(t\mid \mathcal{X}_{s-})$ with $(h,j)$th entry $\alpha_{hj}(t\mid \mathcal{X}_{s-})$, $j \neq h$, and $-\sum_{j\neq h} \alpha_{hj}(t\mid \mathcal{X}_{s-})$ for the $h$th diagonal element.
The \textit{cumulative transition intensities} are given by \[ A_{hj}(t\mid \mathcal{X}_{s-}) := \int_0^t \alpha_{hj}(u\mid \mathcal{X}_{s-}) du\] and can be organised into a $k\times k$ \textit{cumulative transition intensity matrix} $\mathbf{A}(t\mid \mathcal{X}_{s-}) = (A_{hj}(t\mid \mathcal{X}_{s-}))_{h,j\in\mathcal S}$. 

\subsection{Dependency on the History}
\label{sec:dep.hist}

 Future state transitions can depend on the history $\mathcal{X}_{s-}$ in different ways. The simplest form is given by the Markov property. 
A multi-state model is termed a \textit{Markov model},
if all relevant information about its history is solely determined by the current state occupied at time $s$, rather than the specific path taken to reach that state. In this context, 'relevant' signifies the information necessary for predicting future transitions of the process. 
This property implies
\begin{align*}
P_{hj}(s,t) := \mathbb{P}\Bigl(X(t) = j| X(s) = h\Bigr) = P_{hj}(s,t\mid\mathcal{X}_{s-}) .
\end{align*}
It means that the past and future are independent given the present at time $s$
\citep[pp.~462-463]{aalen2008survival}. 


So called Markov extension models additionally include the time spent in the current state as further information, but not the history before entering the state. In this work, \textit{semi-Markov models} are used as a special case of these models. Since they are Markov extension models they do include the duration but they no longer depend on the current time \citep[p.~245]{hougaard1999multi}.

As mentioned before, the Markov assumption may be unsuitable for a number of applications. Then, \textit{non-Markov models} are an alternative, where the transition probabilities between states can depend on the entire history of the process including past states and transitions. 
Different ways how this dependency can be modelled will be introduced in Section \ref{sec:sim_sce}.

It is also plausible to consider a hybrid approach within the same model. Some transitions may adhere to the Markov principle, while others exhibit non-Markov characteristics. This blending of methodologies is termed as \textit{partially non-Markov models} by \citet{maltzahn2021hybrid} and offer flexibility in capturing different types of dependencies within the data.

\subsection{Observations and Censoring}
\label{sec:obs}

The typical observation plan consists of monitoring a set of $n$ subjects with i.i.d. processes $\bigl(X_i(t), ~t \in \mathcal{T}\bigr),~ i = 1,...,n,$ over a specified time period $\mathcal{T} = [0, \tau)$. 
The subjects may start in different states. 

A common feature of event history analysis are incomplete observations.
 There are several reasons why the data is incomplete. One reason is that by the end of the observation period at time point $\tau$, some subjects might not have reached an absorbing state. 
Another possibility is that a subject is lost to follow-up and cannot be observed past a certain time point. 
This is termed \textit{right-censoring}. Here, the precise transition time is unknown,
only that it is greater than a certain value.
 Denote the right-censoring time for the $i$th individual with $C_i$. 
The independence of $C_i$ and $X_i(.)$ is necessary in order to allow for valid inference for the complete data. 
The process 
\[
H_i(t) := \mathbbm{1}\{C_i \geq t\}
\]
indicates if the process of the $i$th subject is observed at time $t$.
The censored version of the multi-state process is denoted by $\tilde{X}_i(t) = X_i(t \wedge C_i)$, with $x\wedge y = \min(x,y)$. 
The observable data \[\Bigl\{\tilde{X}_i(t),~H_i(t),~ t \in [0, \tau] \Bigr\}, ~i= 1,...,n,\] consists of the censored version of the multi-state process and the information about censoring for all time points within the study interval for each subject.
The censored versions of the counting and at-risk processes
are 

\begin{align}
\tilde{N}_{hj}^i(t) &:= \sum_{u\in(0,t]} \mathbbm{1}\Bigl\{\tilde{X}_i(u-)=h,~ \tilde{X}_i(u) := j, ~H_i(u) = 1 \Bigr\}, \notag\\
\tilde{Y}_h^i(t) &:= \mathbbm{1}\Bigl\{\tilde{X}_i(t-) = h,~ H_i(t) = 1\Bigr\},\notag\\
\label{eq:counting_sum}
\tilde{N}_{hj}(t) &:= \sum_{i=1}^n \tilde{N}_{hj}^i(t),\\
\label{eq:risk_sum}
\tilde{Y}_h(t) &:= \sum_{i=1}^n \tilde{Y}_h^i(t)
\end{align}
for all $t\leq \tau$ \citep{glidden2002robust}.

\subsection{Estimators}
\label{sec:est}

There are multiple estimators for the transition probabilities in multi-state models. Under the Markov assumption, the Aalen-Johansen estimator by \citet{aalen1978empirical} provides a consistent estimation.
For non-Markov multi-state models, various estimators have been proposed. Some examples are \citet{de2015nonparametric}, \citet{titman2015transition} and \citet{allignol2014competing}. They are not considered in this work since they are limited to a specific multi-state model, need to fulfil additional conditions or the estimators used in this work showed better results in previous comparisons, for example in \citet{spitoni2018non} or \citet{maltzahn2021hybrid}.
In the following, the Aalen-Johansen, landmark Aalen-Johansen and different versions of the hybrid Landmark Aalen-Johansen estimator are investigated. These estimators are applicable to general multi-state models.

\subsubsection{Aalen-Johansen Estimator}
\label{sec: AJ}

Let $J_h(u) = \mathbbm\{\tilde{Y}_h(u) >0\}$ denote whether any subject is in state $h$ at time $u$. 
Then, the $(h,j)$th element ($h\neq j$) of the the non-parametric \textit{Nelson-Aalen estimator} $\mathbf{\hat{A}}(t)$ is denoted by
\begin{align}
\label{eq:Nel-Aa}
\hat{A}_{hj}(t) &:= \int_0^t \frac{J_h(u)}{\tilde{Y}_h(u)} d \tilde{N}_{hj}(u), ~ h \neq j, 
\end{align}
and diagonal elements are $\hat{A}_{hh}(t) = - \sum_{j \neq h} \hat{A}_{hj}(t)$. 
The \textit{Aalen-Johansen} (AJ) estimator for the transition probability results in
\begin{align*}
\hat{\mathbf{P}}^{\text{AJ}}(s, t) = \prod_{s<t_K\leq t} \bigl(\mathbf{I} + d \hat{\mathbf{A}}(t_K)\bigr)
\end{align*}
\cite[p.~123]{aalen2008survival}.


\subsubsection{Landmark Aalen-Johansen Estimator}

The limitation of the AJ estimator lies in its consistency only within Markov models, potentially resulting in bias in non-Markov scenarios. To address this limitation, \citet{spitoni2018non} proposed the \textit{landmark Aalen-Johansen} (LMAJ) estimator building on the results of \citet{datta2001validity}. It offers a solution tailored for general multi-state models, including non-Markov models. The authors demonstrated its consistency for estimating transition probabilities within such models. This method involves subsetting the data based on landmarking, meaning it only includes subjects that fulfil the requirement of being in a given state - the \textit{landmark state }- at a given time - the \textit{landmark time}. 
Let $s$ be the fixed landmark time point and $h$ the fixed landmark state. Then, the subsample includes all subjects $i$ with $\tilde{X}_i(s) = h$. 
 The superscript $(LM)$ denotes the landmark-based version of the processes.
Hence, $\tilde{N}_{hj}(t)$, $\tilde{Y}_h(t)$, $J_h(t)$ and $\hat{A}_{hj}(t)$ are adjusted to
\begin{align*}
\tilde{N}_{hj}^{(LM)}(t) &:= \sum_{i=1}^n \tilde{N}_{hj}^i(t) \mathbbm{1}\bigl\{\tilde{X}_i(s) = h\bigr\}\\
\tilde{Y}_{h}^{(LM)}(t) &:= \sum_{i=1}^n \tilde{Y}_{h}^i(t) \mathbbm{1}\bigl\{\tilde{X}_i(s) =h\bigr\}\\
J_h^{(LM)}(t) &:= \mathbbm{1}\bigl\{\tilde{Y}^{(LM)}_h(t) >0\bigr\}\\
\hat{A}_{hj}^{(LM)}(t) &:= \int_0^t \frac{J_h^{(LM)}(u)}{\tilde{Y}^{(LM)}_h(u)} d \tilde{N}^{(LM)}_{hj}(u).
\end{align*}
 The fixed $s$ is suppressed in the notation.
Then, the LMAJ estimator is denoted by
\begin{align*}
\hat{\mathbf{P}}_{h}^{\text{LMAJ}}(s, t) :=  \mathbf{e}_h\Prodi_s^t \bigl(\mathbf{I} + d \mathbf{\hat{A}}^{(LM)}(u)\bigr),
\end{align*}
where the entries of $\mathbf{\hat{A}}^{(LM)}$ are ordered as before using $\hat{A}^{(LM)}_{hj}(t)$, and $\mathbf{e}_h$ is a row vector with the $h$th entry equal to $1$ and all other elements $0$.  
It is an estimator for  $\mathbf{P}_h(s, t) =  \Bigl(P_{h1}(s, t),...,P_{hk}(s, t)\Bigr)$.
Thus, the landmark state corresponds to the state out of which the transition happens, and the landmark time is the starting time of the transition probability
\citep{spitoni2018non}.

\subsubsection{Hybrid Landmark Aalen-Johansen Estimator}

The issue with the LMAJ estimator is that it strictly reduces data without checking if the transition is really non-Markov. This would be an unnecessary loss of data in a Markov case and is especially problematic for transitions for which already only a few observations are available. Usually in real life it is not known in advance whether the Markov assumption holds for a particular dataset.
Due to the decreased sample size, the variance might increase. To solve the problem, \citet{maltzahn2021hybrid} proposed the \textit{hybrid landmark Aalen-Johansen} (HAJ) estimator as a compromise between the AJ and the LMAJ estimator. The idea is to improve the estimation precision relative to the LMAJ estimator at the possible expense of introducing bias.

In a first step, for each transition separately, it is determined whether the Markov assumption holds. If it does, all subjects are used for the estimation for that transition. Otherwise, landmarking is applied. The intention behind using the HAJ estimator is to potentially lower the variance compared to the LMAJ estimator due to possibly less reduction in sample size and achieve a smaller bias relative to the AJ estimator by incorporating landmarking for non-Markov behaviour.
If only a subset of all transitions is non-Markov, the model is called partially non-Markov model. This estimator should work especially well for models of this type.

In this work, two versions of this estimator are applied. The first one utilises the log-rank-based test by \citet{titman2022general}, as introduced in Section~A in the supplement. \citet{maltzahn2021hybrid} also based their HAJ estimator on a log-rank test. 
Additionally, a new version is considered, using the Cox model as outlined in Section~\ref{sec:cox.mod}. This decision is motivated by the demonstrated high power of this test across diverse scenarios \citep{titman2022general}. 
 
The methods vary in their approach to non-Markov behaviour: the log-rank-based test includes the condition of being in state $l$ at time $s$, while the Cox model incorporates the most recent time of entering the current state. Thus, different parts of the history are used. Since the testing serves as a diagnostic tool rather than for formal statistical inference, the p-values remain unadjusted for the different transitions within the same model.

After the testing, the procedure follows a similar pattern to the two previous estimators.
Let $\mathcal{M}$ denote the set of non-Markov transitions, determined by either of the two tests.
Then, the estimator for the $(h,j)$th element of the cumulative transition intensity matrix $\mathbf{\hat{A}}^{(H)}(t)$ is defined as
\[
\hat{A}^{(H)}_{hj}(t) := \begin{cases}
\hat{A}_{hj}(t) &,hj \notin \mathcal{M}\\
\hat{A}_{hj}^{(LM)}(t) &, hj \in \mathcal{M}.
\end{cases}
\]
The HAJ estimator is presented by 
\begin{align}
\label{eq: HAJ}
\mathbf{\hat{P}}_h^{\text{HAJ}}(s, t) :=  \mathbf{e}_h\Prodi_{s}^t\bigl(\mathbf{I} + \mathbf{\hat{A}}^{(H)}(du)\bigr).
\end{align}
\citet{maltzahn2021hybrid} showed that the estimator is consistent.
The theory, however, also holds for a set of states from the state space $\mathcal{S}$ \citep{maltzahn2021hybrid, spitoni2018non}.

\subsubsection{Variance}
To estimate the variance for all estimated transition probabilities, the \textit{Greenwood type estimator} is used.
First, the Greenwood type covariance estimator for $d\mathbf{\hat{A}}(t)$ is defined as
\[
\widehat{\text{Cov}}\Bigl(d\hat{A}_{hj}(t), d\hat{A}_{hj'}(t)\Bigr):= \frac{\Bigl(\mathbbm{1}\bigl\{j = j' \bigr\} \tilde{Y}_h(t)  - d\tilde{N}_{hj}(t)\Bigr)d\tilde{N}_{hj'}(t) }{\tilde{Y}_h^3(t)},~\text{if}~h \neq j,~ h \neq j',
\]
and
\[
\widehat{\text{Cov}}\Bigl(d\hat{A}_{hj}(t), d\hat{A}_{h'j'}(t)\Bigr):= 0,~ \text{if}~ h \neq h'.
\]

Then, the variance for the estimated transition probability matrix can be estimated by
\begin{align}
\label{eq:green}
\widehat{\text{Var}}&\Bigl(\mathbf{\hat{P}}^{\text{AJ}}(s,t)\Bigr) := \notag\\ &\sum_{u \in (s, t]} \Bigl \{ 
\mathbf{\hat{P}}^{\text{AJ}}(s,t)^\intercal \otimes \mathbf{\hat{P}}^{\text{AJ}}(s,u-) \Bigr\} \widehat{\text{Cov}}\Bigl(d\mathbf{\hat{A}}(u)\Bigr) \Bigl\{ 
\mathbf{\hat{P}}^{\text{AJ}}(s,t) \otimes \mathbf{\hat{P}}^{\text{AJ}}(s,u-)^\intercal \Bigr\} ,
\end{align} 
where $\otimes$ denotes the Kronecker product. Additionally, a recursion formula is available for computational purposes. For more details, refer to \citet{mstate}.
For the LMAJ and the HAJ estimators, the notation is adjusted accordingly \citep{maltzahn2021hybrid,spitoni2018non}. 


\section{Testing the Markov Assumption}
\label{sec:test}

To test whether the Markov assumption holds true in a multi-state model, different tests are available. In the following, an approach using a Cox proportional hazards model is explained. Furthermore, a log-rank-based statistic developed by \citet{titman2022general} is introduced.

\subsection{Cox Model}
\label{sec:cox.mod}


This method tests deviation from Markov behaviour by considering the most recent entering time into state $h$, denoted by $t_h$, as a covariate in a \textit{Cox proportional hazards model}. 
Let $\theta_{hj}$ be the parameter for transition $h \rightarrow j$. Each transition is modelled separately only using the corresponding data.
Then, the semi-parametric model has the form \[\alpha_{hj}(t\mid \mathcal{X}_t) = \alpha_{hj}(t, t_{h}) =  \alpha_{hj0}(t) \text{exp}(\theta_{hj}t_{h}),\] 
where $\alpha_{hj0}(t)$ is the unspecified non-negative baseline transition intensity. Including additional or other covariates would be possible but is not further investigated in the following.

Let the observed transition times into state $h$ for those processes, that will later move directly to state $j$, be denoted by $t_{h}^1 < t_{h}^2<...<t_{h}^{\tilde{N}_{hj}(\tau)}$. Assume no tied events, meaning that there is at most one transition per time point. 
Define a set of subjects that are at risk at time point $t_{h}^q$, $q \in \bigl\{1,...,\tilde{N}_{hj}(\tau)\bigr\}$ as $\mathcal{R}_{q} = \bigl\{i: \tilde{Y}_i(t_{h}^q) = 1 \bigr\}.$ In order to estimate the parametric component of the model, the partial log-likelihood function is needed given by
 \[
\log\Bigl( L\bigl(\theta_{hj}\bigr)\Bigr) := \log\Biggl(\prod_{q = 1}^{\tilde{N}_{hj}(\tau)}  \frac{\text{exp}(\theta_{hj}t_{h}^q)}{\sum_{p\in \mathcal{R}_{q}} \text{exp}(\theta_{hj}t_{h}^p)}\Biggr). \]
 Then, the estimated parameter is denoted by
 \[\hat{\theta}_{hj}: = \underset{\theta_{hj}}{\text{argmax}}~\log\Bigl(L\bigl(\theta_{hj}\bigr)\Bigr).\]
 This can be solved by using the Newton-Raphson algorithm  \cite[pp.~39-41]{therneau2000cox}. 
 
 In order to determine if the Markov property holds, \[H_0: \theta_{hj} = 0~\text{vs}~H_1: \theta_{hj} \neq 0\] is tested using a \textit{Wald test}.
 The Wald test statistic is denoted by 
 \[
W := \frac{\hat{\theta}_{hj}}{\sqrt{\widehat{\text{Var}}(\hat{\theta}_{hj})}}.
 \] 
 An estimation of the variance is obtained using the inverse of the observed information matrix. For more details refer to \citet{therneau2000cox}.
 Under the null hypothesis, the Wald statistic is approximately standard normally distributed.
If the null hypothesis is rejected, i.e. $|W|> z_{1-\alpha/2}$ with $z_{1-\alpha/2}$ being the $1-\frac{\alpha}{2}$ quantile of the standard normal distribution, non-Markov behaviour is assumed since a part of the history, namely the most recent time entering the state, have a significant influence on the transition intensity. 
Note that here, $\alpha$ represents the significance level rather than the transition intensity \cite[pp.~135-136]{aalen2008survival}.
There are possible extensions, such as using the first time of entrance into state $h$ or the number of previous visits to state $h$, for other forms of deviation from the Markov property \citep{titman2022general}.

\subsection{Log-Rank-Based Test}
\label{sec:tit.test}

\citet{titman2022general} introduced a novel approach to test the Markov assumption in multi-state models, using families of log-rank statistics. The core idea is that under the Markov assumption, transitions occurring at time $t>s$ should not be influenced by the state occupied at time $s$.
Thus, this method test departures from the Markov assumption of this specific form. According to the authors, this test is most useful for cases where the non-Markov behaviour does not persist or is not consistent across time.
First, a local test is introduced, considering the transition $h \rightarrow j$ and groups of subjects in a specific state $l$ at a specific time $s$. This local test is then extended to a global test, using a range of time points $s$ for a specific state $l$. In a third step, the approach is broadened to include multiple qualifying states $l$ for a transition specific statistic. Lastly, the asymptotic null distribution for that statistic is determined by using a wild bootstrap. 
Details on the log-rank-based test can be found in Section~A in the supplement.

\section{Data Generation and Analysis Plan}
\label{sec:data}


To compare the performance of the previously introduced estimators under controlled conditions, a simulation study is conducted under a range of settings. The study focusses on a three-state illness-death model with recovery, as depicted in Figure \ref{fig:illness_death}. 

\subsection{Parametric Distribution for Event History Data and Censoring} 
\label{sec:param dist}

Different distributions are used to model continuous event history data. A common one is the Weibull distribution, characterised by a scale parameter $a_{hj} > 0$ and a shape parameter $b_{hj} > 0$. Thus, each transition time follows $T_{hj} \sim \text{Weibull}(a_{hj}, b_{hj})$. 

To acquire realistic parameter values for the illness-death model with recovery in the simulation, the \texttt{prothr} dataset from the {R} package \texttt{mstate} is used \citep{mstate}. The dataset contains information about patients derived from a randomised clinical trial focusing on prednisone treatment in liver cirrhosis. 
Patient recruitment occurred between 1962 and 1969, with the final follow-up date set at September 1974, resulting in a maximum study time of $4892$ days and $n = 488$ subjects.
The different states the subjects can occupy are: 
\begin{itemize}
\item State 1: low prothrombin level
\item  State 2: high prothrombin level
\item  State 3: death
\end{itemize}
Since recovery from death is not possible, but an exchange between high and low prothrombin levels is feasible, the illness-death model is suitable for this data.
Overall, the time of entry and exit in days, the transition from and to which state and censoring are considered. Each patient's time starts at $0$.

%

In order to obtain parameters for the four possible transitions in that dataset, a Weibull model is fitted via the function \texttt{flexsurvreg} from the {R} package \texttt{flexsurv} \citep{flexsurv}. Each transition is modelled separately. 
 Notably, given the possibility of recovery, subjects may be included multiple times for both the $1\rightarrow 2$ and the $2 \rightarrow 1$ transition.

For notational simplicity, the transition subscript $hj$ will be omitted. 
Let $t_{i*}$, with $i*=1,...,\tilde{N}$, 
 be the times of transitions of interest, where $\tilde{N}$ is the total number of transitions out of state $h$ in the dataset. 
 Let $c_{i*}$ be the censoring indicator, meaning if $t_{i*}$ is an observed transition $h \rightarrow j$, then $c_{i*} = 0$.
  Otherwise, it is $c_{i*} = 1$. Let $w_{i*}$ be the left-truncation time if applicable. 
If there is no left-truncation, then $w_{i*} = 0$. Since each transition is considered separately, the data for the $2\rightarrow 1$ and $2 \rightarrow 3$ transitions for a subject is possibly left-truncated if this subject is previously in state~$1$ before moving to state $2$. The same applies analogously for transitions $1 \rightarrow 2$ and $1 \rightarrow3$ with occupying state $2$. This type of truncation, where there is only a delayed entry into a certain state, but the subject was already observed previously in another state is called \textit{internal left truncation} \citep[p.~175]{beyersmannR}.
The censoring times are assumed to be independent of the transition times. Additionally, the individual transition times are assumed to be independent. It should be noted that this assumption might not hold, particularly given that multiple transitions from the same subject can enter the dataset for estimation, and the Markov assumption may not necessarily hold for the original data. However, since the primary aim is to obtain realistic parameters for a simulation, further investigation into this assumption is not pursued.

 To fit a fully-parametric model, the log-likelihood function 
 is maximised using the BFGS algorithm by \citet{nash1990}. 

To acquire estimations for parameters representing a semi-Markov process, the entry and exit times are modified. All entry times are reset to $0$ and the exit times are modified to be the sojourn time in the respective state by subtracting the original entry time from the original exit time. 
Otherwise, the same procedure is applied as before.

The resulting parameters are displayed in Table \ref{tab:param}. 
The transition $2 \rightarrow 1$ in the semi-Markov model is the only transition with an increasing intensity over time. This is due to the estimated shape parameter $\hat{b}_{21} > 1$. All other parameters indicate a decreasing intensity with shape parameters smaller than 1. 

\begin{table}[t]
\caption{Estimated Weibull parameters for the transition intensities for a Markov and semi-Markov model in an illness-death model with recovery using the \texttt{prothr} dataset. 
}
\centering
\begin{tabular}{|c|c|c|}
\hline
Estimated Parameters & Markov & semi-Markov \\
\hline
$(\hat{a}_{12}, \hat{b}_{12})$ & $(0.0057, 0.7050)$ & $(0.0037, 0.7453)$ \\
\hline
$(\hat{a}_{13}, \hat{b}_{13})$ & $(0.0003, 0.9449)$ & $(0.0006, 0.8682)$ \\
\hline
$(\hat{a}_{21}, \hat{b}_{21})$ & $(0.0058, 0.8327)$ & $(0.0015, 1.0228)$ \\
\hline
$(\hat{a}_{23}, \hat{b}_{23})$ & $(0.0017, 0.9253)$ & $(0.0046, 0.7549)$ \\
\hline
\end{tabular}
\label{tab:param}
\end{table}

Before giving more details about the utilisation of the estimated parameters in the different settings, the procedure of drawing censoring times is explained.
For each subject a separate right-censoring time is determined.
First, an exponential distributed random variable with parameter $\lambda = 0.00035$ is drawn, $c_{i1} \sim \text{Exp}(\lambda)$. If $ c_{i1} > 2500$, a uniform random variable $c_{i2} \sim U(2500, 5200)$ is drawn. 
Furthermore, the administrative right-censoring time $\tau = 4892$ represents the end of study for all subjects as in the original data. Then, the censoring time $c_i$ is either $\min\{c_{i1}, \tau\}$ or $\min\{ c_{i2}, \tau\}$, whichever is applicable.
The parameters are chosen, such that the right-censoring proportion and distribution resemble that of the original data.


\subsection{Simulation Settings}
\label{sec:sim_sce}

In the following, six different settings are presented for the simulation study. 
Each setting will be introduced in detail after the general simulation set up is explained. The following details are chosen to resemble the original data as closely as possible.
There are $10000$ simulation runs per setting. One run contains $n = 488$ individual multi-state trajectories. 
For each subject, a censoring time $c_i$ is drawn as previously described. 
  The starting state $1$ or $2$ for each subject is randomly assigned, with a probability of $44.67\%$ for state $1$ and $55.33\%$ for state $2$. 
%
%
%

\tikzstyle{block} = [rectangle, draw, 
fill = lightgray,
draw = lightgray,
    text width=25em, 
    text centered, rounded corners, minimum height=2.2em 
    ]
\tikzstyle{blocka} = [rectangle, draw, 
fill = white,
draw = white,
    text width=4em, 
    text centered, rounded corners, minimum height=2.8em 
    ]

\begin{figure}[t]
 \centering
\scalebox{0.8}{
 \begin{tikzpicture}[node distance = 1.8cm, auto]

 \node [block] (pII) {$h$: current state\\$\mathcal{R}_h$: set of states reachable from $h$};  
 \node[blocka, below of =  pII, text width = 4em](pL){};
   \node [block, left of=pL,node distance=2.7cm,text width=11em] (pIV) {$\forall j \in \mathcal{R}_h$ use $\alpha_{hj}:$ $t_{hj}$}; 
     \node [block,   right of=pL,node distance=2.7cm,text width=11em] (pE) {stop}; 
    \node [block, below of=pIV, text width=11em] (pV) {$t^* = \min\{t_{hj_1} , t_{hj_2} , c_i\}$}; 
    \node[blocka, below of =  pE, text width = 1 em](pLi){$~~~$};
    \node[block, below of =  pV, text width = 11 em](pVI){set $h  ={\textnormal{argmin}_{j\in \mathcal{R}_h}t_{hj}}$};
        \node[block, below of =  pLi, text width = 11 em](pVII){ stop};
	
	\draw[->] (pII) -- node[left = 4pt] {$\mathcal{R}_h \neq \emptyset$} (pIV) ;
		\draw[->] (pV) -- node[left = 4pt] {$t^* \neq c_i$} (pVI) ;
\draw[->] (pV.south east) -- node[right=6pt] {$t^* = c_i$} (pVII);
	\draw[->] (pII) -- node[right= 4pt] {$\mathcal{R}_h = \emptyset$} (pE) ;

   \draw [->] (pV) -- (pVI);
     \draw [->] (pIV) -- (pV);  
     \draw [->] (pVI) to [out=180,in=180] (pII);
	
 \end{tikzpicture}
 }
 \caption{Flowchart for generating new event times $t^*$ and corresponding new event types $j^*$ for each subject in the simulation study.}
  \label{fig:scheme}
  \end{figure}
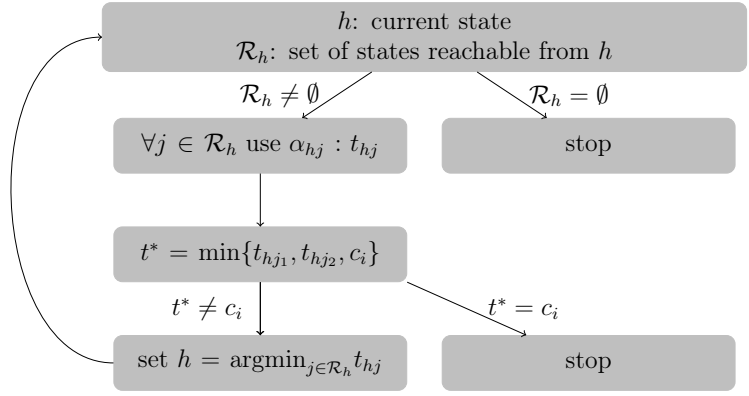

The transition times and types of events for each subject are determined by following the steps outlined in Figure \ref{fig:scheme}. Initially, the current state $h$ is set to the specified starting state and the set $\mathcal{R}_h$ of reachable states from $h$ is determined. For instance, if the starting state is state 1, there are two reachable states: state~2 and state~3, denoted by $\mathcal{R}_1 = \{2,3\}$. In the illness-death model with recovery, there are always two possible transitions if the subject is not in the absorbing state 3. For each potential transition, a transition time $t_{hj}$ is calculated using the transition intensity~$\alpha_{hj}$ with the 
parameters from Table \ref{tab:param}. The specific method for deriving these transition times varies depending on the settings, which is explained in detail when the settings are introduced. Next, the two transition times and the censoring time are compared. If the censoring time is found to be the lowest, the generation of this sample path is completed. Otherwise, the current state~$h$ is updated to the state associated with the smallest transition time, and the entire process is repeated. If the transition to state 3 has the smallest value, resulting in $\mathcal{R}_3 = \emptyset$, the procedure halts. Alternatively, if the value for state 2 is smaller, leading to $\mathcal{R}_2 = \{1,3\}$, the process repeats as before. Thus, the generation of a sample path continues until either state 3 is reached or the censoring time is the lowest.

\subsubsection{Setting 1: Markov Model}
\label{sec: markov}
The first setting is a time-inhomogeneous Markov process. 
To generate the transition times~$t_{hj}$, the following procedure is implemented. Each potential transition is considered individually, aligning with the classical survival model. 

The procedure begins by drawing a consecutive event time $t^*$ from the conditional distribution given that the previous event occurred at time $t_{h,0}$. 
For the first transition time, it is $t_{h,0} = 0$. For the later transitions, it is $t_{h,0} = t^*$, meaning it is set to the time point of the previous transition. 
In particular, in the Markov model with Weibull-based transition hazards,
the next transition time is generated as \begin{equation*}
\label{eq:trans_time}
\begin{aligned}
 t_{hj} =  \Bigl(t_{h,0}^{\hat{b}_{hj}}-\frac{1}{\hat{a}_{hj}} \log(u_{hj})\Bigr)^{\frac{1}{\hat{b}_{hj}}}.
\end{aligned}
\end{equation*} 
by using the corresponding parameter values for the Markov model taken from Table \ref{tab:param} and drawing $u_{hj}$ from the $\mathcal{U}[0,1]$ distribution, cf. Section~B in the supplement for details.


\subsubsection{Setting 2: Semi-Markov Model}
\label{sec:semi}

The simulation of the transition time for a semi-Markov process is similar to the one described in the previous section. The difference here is, that only the sojourn time in the current state is relevant. Therefore, $t_{h,0}$ is not needed in Equation (\ref{eq:trans_time}). Instead, times are generated as 
\begin{align*}
\tilde{t}_{hj} = \Bigl(-\frac{1}{\hat{a}_{hj}} \log(u)\Bigr)^{\frac{1}{\hat{\vphantom{b}\smash[b]{b}}^{}_{hj}}},
\end{align*} 
by sampling $u$ from the uniform distribution.
The total transition time is then given by 
\begin{align}
\label{eq:semi_time}
t_{hj} = t_{h,0} + \tilde{t}_{hj},
\end{align}
 with $t_{h,0} = 0$ for the first transition. If $t^* \neq c_i$, for the next iteration, $t_{h,0}$ is set to the value of (\ref{eq:semi_time}) from the previous iteration.
This way, only the sojourn time in the current state has an influence on generating the time for the next transition, while still keeping track of the total time. 
The rest of the procedure follows as depicted in Figure \ref{fig:scheme}.
For this setting, the semi-Markov parameters from Table \ref{tab:param} are used. 
Those are needed since this form of generating the times is a version of the \textit{clock-reset} approach. To generate the new time $\tilde{t}_{hj}$, the start time is technically reset to 0. In all other settings, the \textit{clock forward} approach is used, meaning that the time since the subject entered the first state is considered to generate the time of the next event \citep{putter2007tutorial}.

\subsubsection{Setting 3: Frailty Model}
\label{sec:non}

In the third setting a frailty model is used to generate non-Markov processes. 
 Here, a subject specific variable $W_i$ is considered to introduce dependency between transitions for the same subject, which violates the Markov assumption. 
Let $\alpha_{hj}(t)$ be the baseline transition intensity of the the Weibull transition intensity, valid for every subject. The parameters for this and following settings are taken from Table \ref{tab:param} using the Markov version.
Then, the transition intensity for subject $i$ is given by \[\tilde{\alpha}_{hj}(t|W_i) := W_i \alpha_{hj}(t),\] 
where $W_i$ is a Gamma distributed random variable with shape parameter $\eta_1$ and scale parameter $\eta_2$.
That way it is possible to model subjects that transition more quickly between states if $W_i > 1$, or slower when $W_i < 1$. For $W_i = 1$ this is setting 1 \citep[p.~95]{van2016multi}.

Analogue to Equation (\ref{eq:trans_time}), the transition times are generated by 
\begin{align}
\label{eq:frail_alpha}
t_{hj} =  \Bigl(t_{h,0}^{\hat{b}_{hj}}-\frac{1}{W_i \hat{a}_{hj} } \log(u)\Bigr)^{\frac{1}{\hat{b}_{hj}}}.
\end{align}

To evaluate the performance of all estimators for different magnitudes of the frailty variable and therefore violation of the Markov assumption, three parameter combinations for the Gamma distribution are considered:
\begin{enumerate}
\item[a)] $\eta_1 = 2$, $\eta_2 = \frac{1}{2}$
\item[b)] $\eta_1 = 1$, $\eta_2 = 1$
\item[c)] $\eta_1 = \frac{1}{2}$, $\eta_2 = 2$
\end{enumerate}
The expected value and variance of the Gamma distribution are 
\begin{align*}
\text{E}(W_i) &= \eta_1\eta_2\\
\text{Var}(W_i) &= \eta_1\eta_2^2,
\end{align*}
which means that the expected value is 1 in all cases \citep[p.~98]{van2016multi}. The variance has values a) 0.5, b) 1, and c) 2, which corresponds to an increasing right-skewed frailty distribution \citep{maltzahn2021hybrid}. 

\subsubsection{Setting 4: Partially Frailty Model}
To introduce a partially non-Markov model, setting 3 is adjusted. For transition~${2 \rightarrow 1}$, Equation (\ref{eq:frail_alpha}) is used to draw times, meaning that this transition is modelled as non-Markov. For all other transitions, Equation (\ref{eq:trans_time}) from setting 1 is applied and, therefore, are considered to be Markov. 
Three versions of this model are employed, each utilising the parameters a), b) and c) as in settings 3 for the non-Markov transition. 

\subsubsection{Setting 5: Mixed Model}
\label{sec:mix}
In the fifth setting, a mixed process is introduced. Initially, a Markov process, as detailed in setting 1, is simulated until time point $607.54$. Subsequently, the time generation changes to the frailty model outlined in setting 3, with the three options for varying frailty variances. As an additional step in each iteration of Figure \ref{fig:scheme}, the drawn time are examined if they exceed time point $607.54$. If that is the case, $t_{h,0}$ is set to $607.54$, and applying Equation (\ref{eq:frail_alpha}), a new time is drawn for that iteration.

The choice of time point $607.54$ corresponds to the time when one-third of all patients have reached state 3 in the simulated data used to calculate the true transition probability for the Markov setting, as described in Section \ref{sec:true}. This ensures, that an adequate number of subjects remains available for the analysis with the new characteristics. One of many potential applications for such a model is to account for the progression of treatment, where initially, all patients may exhibit similar responses, but over time, individual treatment responses or lifestyle factors may begin to influence outcomes.

\subsubsection{Setting 6: Pathological non-Markov Model}
\label{sec:path}
The last setting is a pathological non-Markov process. The transition intensity depends on a condition from the past, here the state occupation at a certain time.
The transition intensity is modified as
\[\tilde{\alpha}_{hj}(t) := 
\begin{cases}
z  \alpha_{hj}(t) &,\text{if} ~X(t) = 1 ~\forall t \leq 40 \\
\alpha_{hj}(t) &,\text{else},
\end{cases}
\]
with $\alpha_{hj}$ being the Weibull transition intensity.
The parameter value $z$ is chosen as $0.3$ for all transitions except $2 \rightarrow 1$, where $z=3$. 
This adjustment can be interpreted as follows: a healthy subject starting and staying in the initial state until time point~40 is less likely to fall ill or die and more likely to recover from the illness.
The values for $z$ are chosen arbitrarily, with the only requirement being that the value for recovery from illness is greater than 1 and the others are less than 1. Additionally, the time point 40 is selected early enough to ensure an adequate number of subjects in both groups of possible transition intensities.


\subsection{Simulating the True Transition Probabilities}
\label{sec:true}

\begin{figure}[th!]
        \centering
        \includegraphics[width=\textwidth]{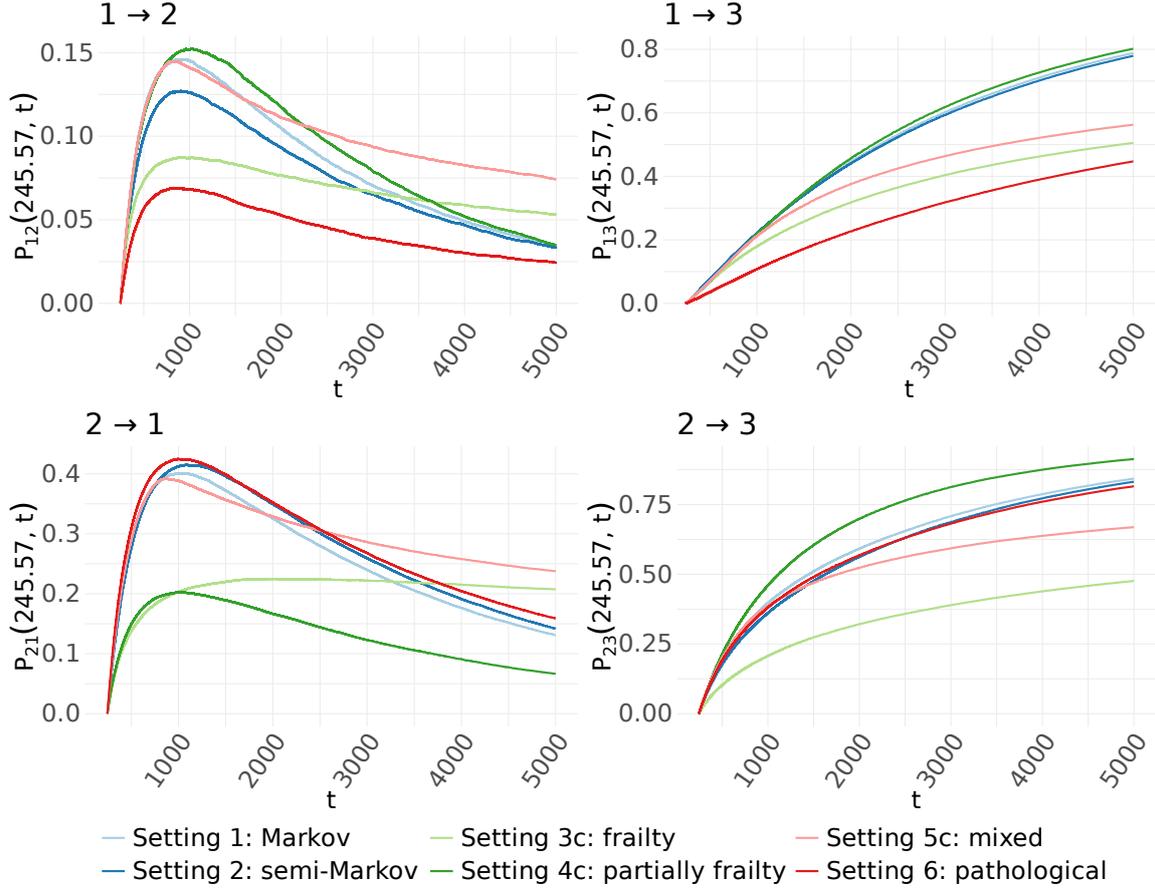}
        \caption{True transition probabilities for the different settings for all transitions. Where applicable, the version of the setting with the highest variance is depicted.} 
        \label{fig:true_p}
\end{figure}

In order to compare the performance of estimators across the different settings, it is necessary to determine the true transition probabilities. Since closed-form solutions only exist for some special multi-state models, another simulation is required for each of the settings \citep{hougaard1999multi}.
To obtain the values, a procedure similar to Figure~\ref{fig:scheme} is applied, with the difference that no censoring occurs. For each setting, $n = 500000$ trajectories are drawn. Transition times for all subjects are drawn as described previously until they reach the absorbing state 3.
The true transition probability is approximated by
\[ 
P_{hj}(s,t) := \frac{\sum_{i=1}^n \mathbbm{1}\bigl\{i: X_i(t) = j, X_i(s) = h\bigr\}}{\sum_{i=1}^n \mathbbm{1}\bigl\{i: X_i(s) = h\bigr\}}
\] 
\citep{maltzahn2021hybrid}.
The true transition probabilities for the different settings and transitions are depicted in Figure \ref{fig:true_p}.

\subsection{Analysis Plan}
\label{sec:crit}

For each setting, the four estimators introduced in Section \ref{sec:est} are calculated for the observable transitions $1 \rightarrow 2$, $1 \rightarrow 3$, $2 \rightarrow 1$ and $2 \rightarrow 3$.
The 0th, 15th and 25th percentile of the time-to-absorption distribution of the simulated ground truth for the Markov setting, introduced in Section \ref{sec:true}, are considered as the different starting times $s$ and therefore also landmark time points. Those are $0$, $245.57$ and $495.64$ and correspond to the time points at which 0\%, 15\% and 25\% of the subjects have reached the absorbing state 3, respectively.
The choice of these time points is purely practical, given the absence of a particular underlying research question for the simulated data. Multiple time points have been chosen to compare the results for varying starting times, with the condition that there are enough observed events for the estimation.
While this practical consideration is relevant, in real applications the choice of starting times may also depend on the research question. 
 
The tests in the hybrid estimators are performed at a 5\% significance level. No adjustment for multiple testing is applied since the tests are primarily seen as diagnostic tools. Within the log-rank-based version, the bootstrap is iterated $n_B = 500$ times. The grid of time points for the testing is chosen out of 60 equally spaced times over the period $[0, 4892]$ with weights from Equation~(A.3) in the supplement greater than 2.5. The exact number of finally used time points depends on the data in each bootstrap iteration since multiple factors play a role as described in Section~A.2 in the supplement.

To evaluate the performance of estimators, \citet{coverage} suggests different measures. In the following, the superscripts AJ, LMAJ, and HAJ are omitted, as the measures are calculated uniformly across all estimators. Thus, $\hat{P}_{hj}$ can represent any of the estimators.
The first measure is the {bias},
which is estimated by
\[
\widehat{\text{Bias}}\Bigl(P_{hj}(s,t), \hat{P}_{hj}(s,t)\Bigr) = \frac{1}{10000}\sum_{g =1}^{10000} \Bigl(\hat{p}_{hj}^{(g)} (s,t) - P_{hj}(s,t)\Bigr),\]
where $\hat{p}_{hj}^{(g)} (s,t)$ corresponds the realisation of the estimated transition probability of the $g$th simulation run, $g = 1,..., 10000$. 
Moreover, the {variance} 
is estimated by
\[
\widehat{\text{Var}} \Bigl(\hat{P}_{hj}(s,t)\Bigr)= \frac{1}{10000 - 1} \sum_{g = 1}^{10000} \Bigl(\hat{p}^{(g)}_{hj}(s,t) - \frac{1}{10000} \sum_{g=1}^{10000} \hat{p}^{(g)}_{hj}(s,t)\Bigr)^2.
\]
Furthermore, the root mean squared error (RMSE)
is estimated by
\[
\widehat{\text{RMSE}}\Bigl(P_{hj}(s,t), \hat{P}_{hj}(s,t)\Bigr) = \sqrt{ \widehat{\text{Var}} \Bigl(\hat{P}_{hj}(s,t)\Bigr) +  \widehat{\text{Bias}}\Bigl(P_{hj}(s,t), \hat{P}_{hj}(s,t)\Bigr)^2}  . 
\]
Lastly, the {pointwise empirical coverage rate} 
states the proportion of confidence intervals that contain the true transition probability.
The estimated confidence interval per simulation run is given by 
\[
\hat{p}_{hj}^{(g)}(s,t) \pm z_{1-\alpha / 2} \sqrt{\widehat{\text{Var}}\Bigl(\hat{p}_{hj}^{(g)}(s,t)\Bigr)},
\] 
with  $z_{1-\alpha / 2}$ being the $(1 - \alpha/2)$-quantile of the standard normal distribution and the estimated variance is given by Equation (\ref{eq:green}). Note that here, $1-\alpha$ represents the confidence level rather than the transition intensity. 
A grid of 800 time points evenly spaced between 0 and 4892 is applied for $t$ to calculate all measures. For later starting times the grid is truncated accordingly.

\subsection{Software}
\label{sec:software}

All calculations are performed using \texttt{R} version 4.5.0 (2025-04-11) \citep{R}.
The \texttt{etm} package along with its corresponding function of the same name, is used to calculate the AJ estimator \citep{etm}.
The \texttt{mstate} package contains the function \texttt{lmaj} to calculate the LMAJ estimator \citep{mstate}.
The \texttt{multistate} package, developed for the original HAJ estimator by \citet{maltzahn2021hybrid}, is used for the calculations of the HAJ$^{\text{LR}}$. Additionally, it is combined with some functions from the \texttt{mstate} package to implement the new version using the Cox model. The model is built using the \texttt{flexsurv} package \citep{flexsurv}.
The log-rank-based test in the HAJ$^{\text{LR}}$ is performed using the code by \citet{titman2022general}, available on \href{https://github.com/andrewtitman/MarkovTest}{GitHub}.

\section{Results}
\label{sec:res}






In this section, the results of the simulation study will be discussed.
Initially, the estimators will be compared based on the transition $1 \rightarrow 2$ with a starting time $245.57$. This aims to provide a general overview of the settings in which each estimator performs well.
To enhance readability, the term `estimated' will be omitted before bias, variance, and RMSE.


\subsection{Comparison of the Bias} 
\begin{figure}[bt]
\centering
\includegraphics[scale=0.3]{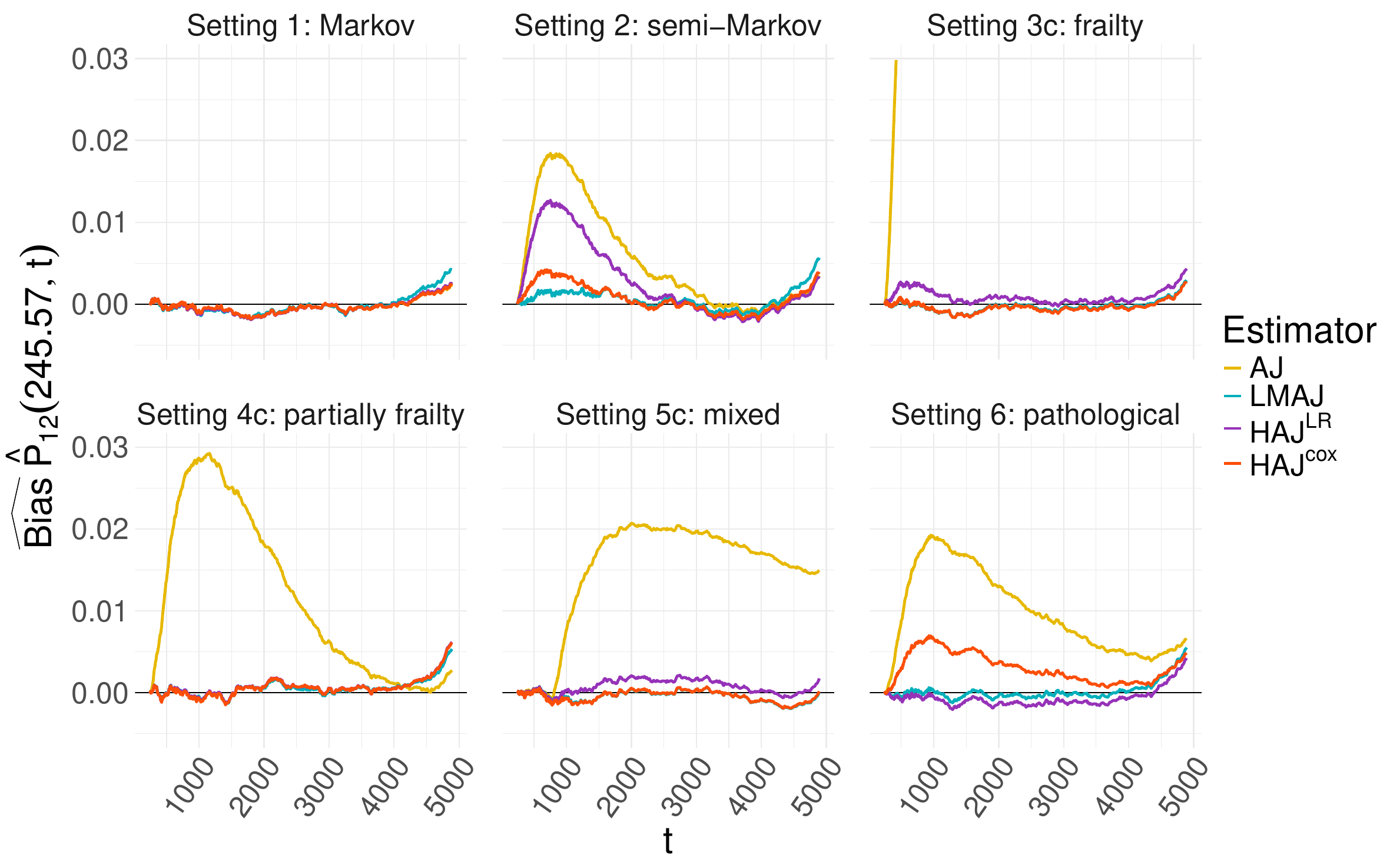} 
\caption{Bias for $\hat{P}_{12}(245.57, t)$ for the settings 1, 2, 3c, 4c, 5c and 6 for the AJ, LMAJ, HAJ$^{\text{LR}}$ and HAJ$^{\text{cox}}$ estimator.}
\label{fig:bias_12_205}
\end{figure}

Figure \ref{fig:bias_12_205} displays the bias for $\hat{P}_{12}(245.57, t)$ for settings 1, 2, 3c, 4c, 5c, and 6, covering the time interval $t\in[245.57, 4892)$.
 This means, where applicable the highest variance for the frailty variable is used,  as the primary focus currently lies on the general comparison of the settings. The impact of different frailty variances on the estimation will be evaluated in detail in Section~C.2 in the supplement.

For the Markov process in setting 1, the bias is small and comparable across all four estimators. However, noticeable differences emerge in the other settings. In settings 2-6, the AJ estimator consistently exhibits the largest bias, particularly pronounced in setting 3c, as shown in Figure~D.1 in the supplement. Here, the y-axis is extended to (0, 0.13) to accommodate the wider range of bias values observed in this setting.
The true transition probabilities for this process take values between 0 and 0.087 depending on the time point $t$, as illustrated in Figure \ref{fig:true_p}.
Therefore, a bias up to 0.130 indicates substantial deviations from the true values. This confirms the expectation, that the AJ estimator is only performing well if the process is Markov and highlights the importance of verifying this assumption, rather than assuming it.

The LMAJ estimator consistently demonstrates the lowest bias close to 0 in all settings, indicating high accuracy.
 In setting 1 and 4c, both HAJ estimators exhibit behaviour similar to the LMAJ.
 Comparing the two HAJ estimators directly reveals, that the Cox version yields lower values in setting 2, 3c and 5c, while the log-rank version has a smaller bias in setting 6. This suggests that the HAJ$^{\text{cox}}$ estimator may offer the intended improvement in those settings with a lower bias.

\subsection{Comparison of the Variance}\label{sec:varcomp} 
\begin{figure}[bt]
\centering
\includegraphics[scale=0.3]{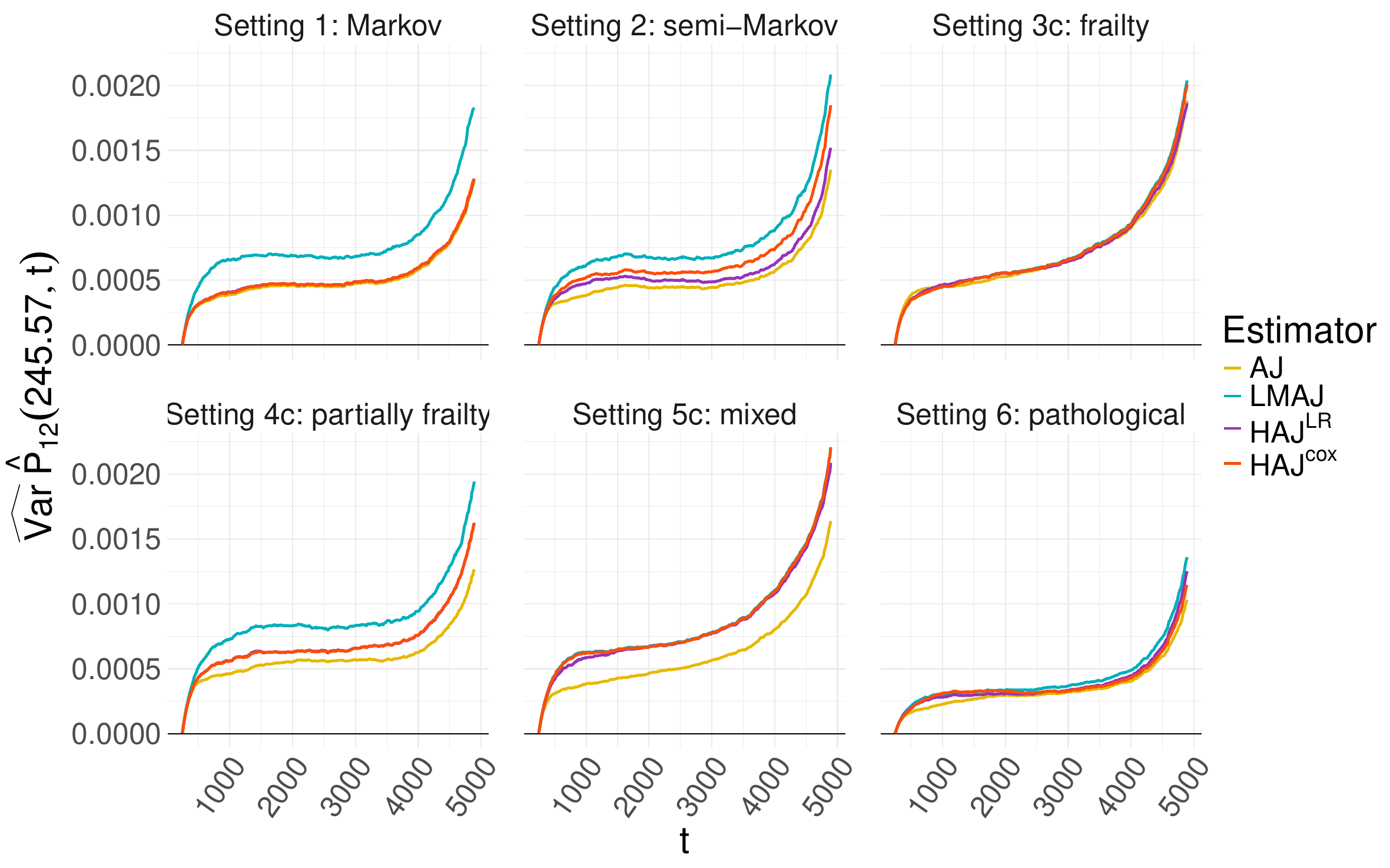}
\caption{Variance for $P_{12}(245.57, t)$ for the settings 1, 2, 3c, 4c, 5c and 6 for the AJ, LMAJ, HAJ$^{\text{LR}}$ and HAJ$^{\text{cox}}$ estimator.}
\label{fig:var_12_205}
\end{figure}

However, when comparing estimators, it is crucial to consider not only the bias but also the variance, as both components contribute to the overall accuracy and reliability of the estimator. 
Therefore, the variance of $\hat{P}_{12}(245.57, t), ~t\in[245.57, 4892)$, for the four estimators is illustrated in Figure \ref{fig:var_12_205}. The AJ estimator always yields amongst the lowest values. This is as expected, given that this method does not involve a reduction in the number of subjects, unlike the other estimators.
Notably, in settings 3c and 6, all estimators exhibit similar variance values. This suggests that, based on the variance alone, there is no discernible advantage for any specific estimator within this setting.

In setting 1, there is almost no performance difference between the HAJ estimators and the AJ estimator, indicating that less landmarking is applied for this hybrid estimator, resulting in similar variance values. 
In contrast for settings 2 and 4c, the values of both HAJ estimators lie between those of the LMAJ and AJ estimator. This demonstrates that both hybrid estimators serve as a compromise, considering more subjects than the LMAJ estimator. Therefore, they yield a more stable estimation. 
In setting 5c, the HAJ estimators exhibit variances similar to those of the LMAJ estimator.

\subsection{Comparison of the RMSE}\label{sec:compRMSE} After the bias and variance have shown partly different results, it is interesting to combine both by considering the RMSE to compare the overall performance of the estimators. 
For setting 3c, this measure is depicted in Figure~D.2 in the supplement.
As expected, due to the comparable variance and large bias, the AJ estimator yields the highest values. Its advantage of a lower variance due to a larger sample size is not applicable in this setting. This means, that it performs considerably worse than the other estimators, which all have similar values.

\begin{figure}[bt]
\centering
\includegraphics[scale=0.3]{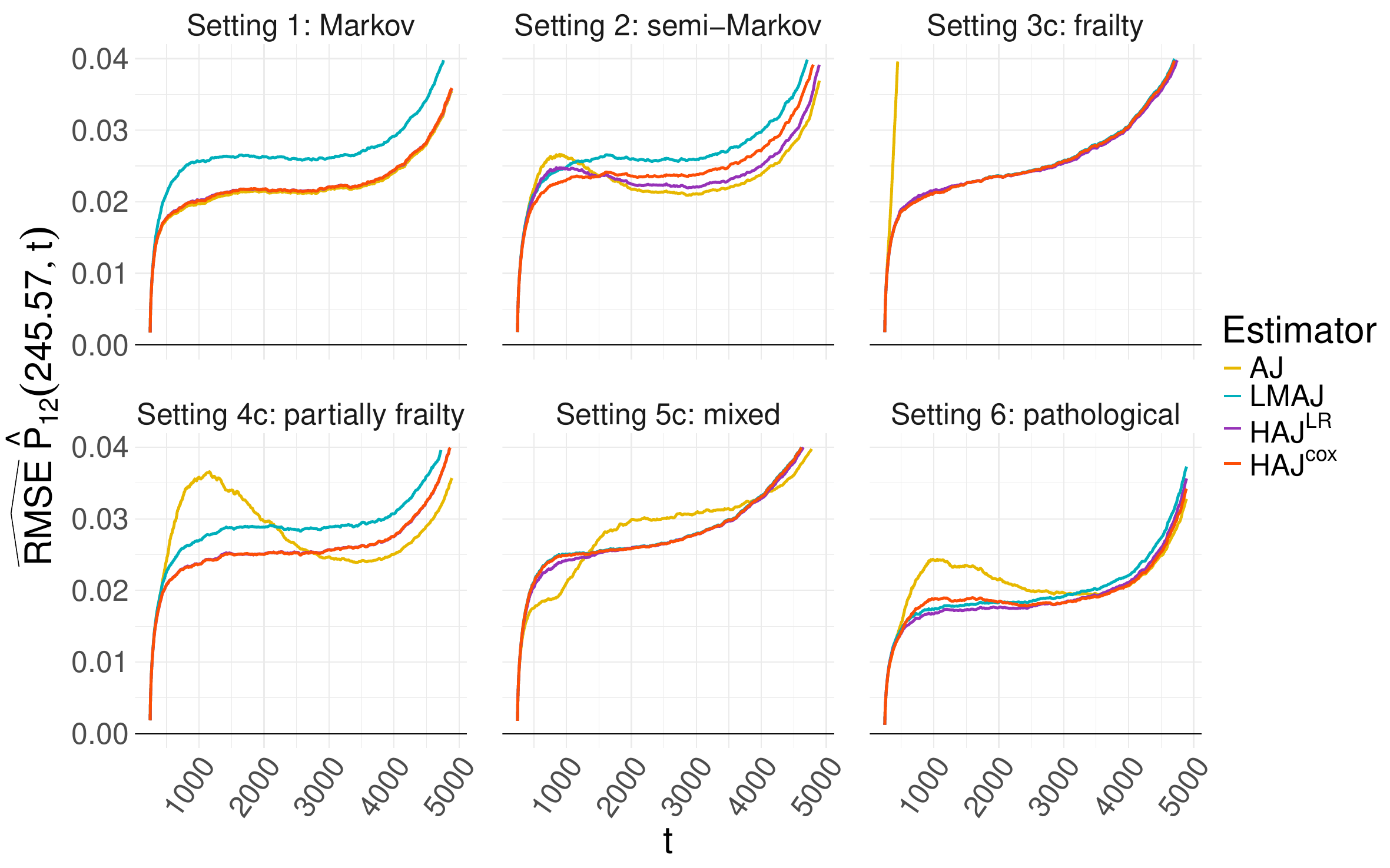}
\caption{RMSE for $P_{12}(245.57, t),$, for the settings 1, 2, 3c, 4c, 5c and 6 for the AJ, LMAJ, HAJ$^{\text{LR}}$ and HAJ$^{\text{cox}}$ estimator.}
\label{fig:rmse_12_205} 
\end{figure}

The RMSE for all settings is depicted in Figure \ref{fig:rmse_12_205}. 
In the Markov setting (setting 1), the AJ estimator and the HAJ estimators yield the lowest RMSE overall, due to their minimal variance and negligible bias. Considering the large values for the AJ estimator in setting 3c and the fact, that in real life it is not known if the Markov assumption holds, the HAJ estimators can therefore be a good alternative.

For the remaining settings, it depends on the time of evaluation $t$, which estimator exhibits the lowest RMSE.
In setting 2, the AJ estimator still has a quite low RMSE, especially for larger time points. The LMAJ and HAJ estimators have a lower RMSE than the AJ estimator for larger time points and a larger RMSE for larger time points. Here, the HAJ$^{\text{LR}}$ estimator performs more similar to the AJ estimator.
In setting 4c, both HAJ estimators show low values overall. Although the AJ estimator has a lower RMSE than the HAJ estimators for later time points, the difference between the estimators is relatively small. The difference for the earlier time points, where the HAJ estimators are better, however, is much larger. So overall, this indicates a preference for the HAJ estimators in this setting.
In setting 5c, the performance of the AJ estimator clearly has a higher RMSE after the change from Markov to non-Markov behaviour at time point 607.54. 
In setting 6, the LMAJ and the HAJ estimators exhibit the lowest RMSE for a wide range of time points $t$.
Notably, the HAJ$^{\text{cox}}$ estimator has a slightly increased RMSE for smaller time points compared to the LMAJ and HAJ$^{\text{LR}}$ estimators.

Overall, in the non-Markov cases 3c-6, the LMAJ and the HAJ estimators demonstrate favourable results, surpassing those of the AJ estimator. However, exceptions arise with the LMAJ estimator in setting 4c. Consequently, the HAJ estimators exhibit the best overall performance, since they additionally show preferable results in settings 1 and 2.

\subsection{Comparison of the Empirical Coverage Rate} 
\begin{figure}[bt]
\centering
\includegraphics[scale=0.3]{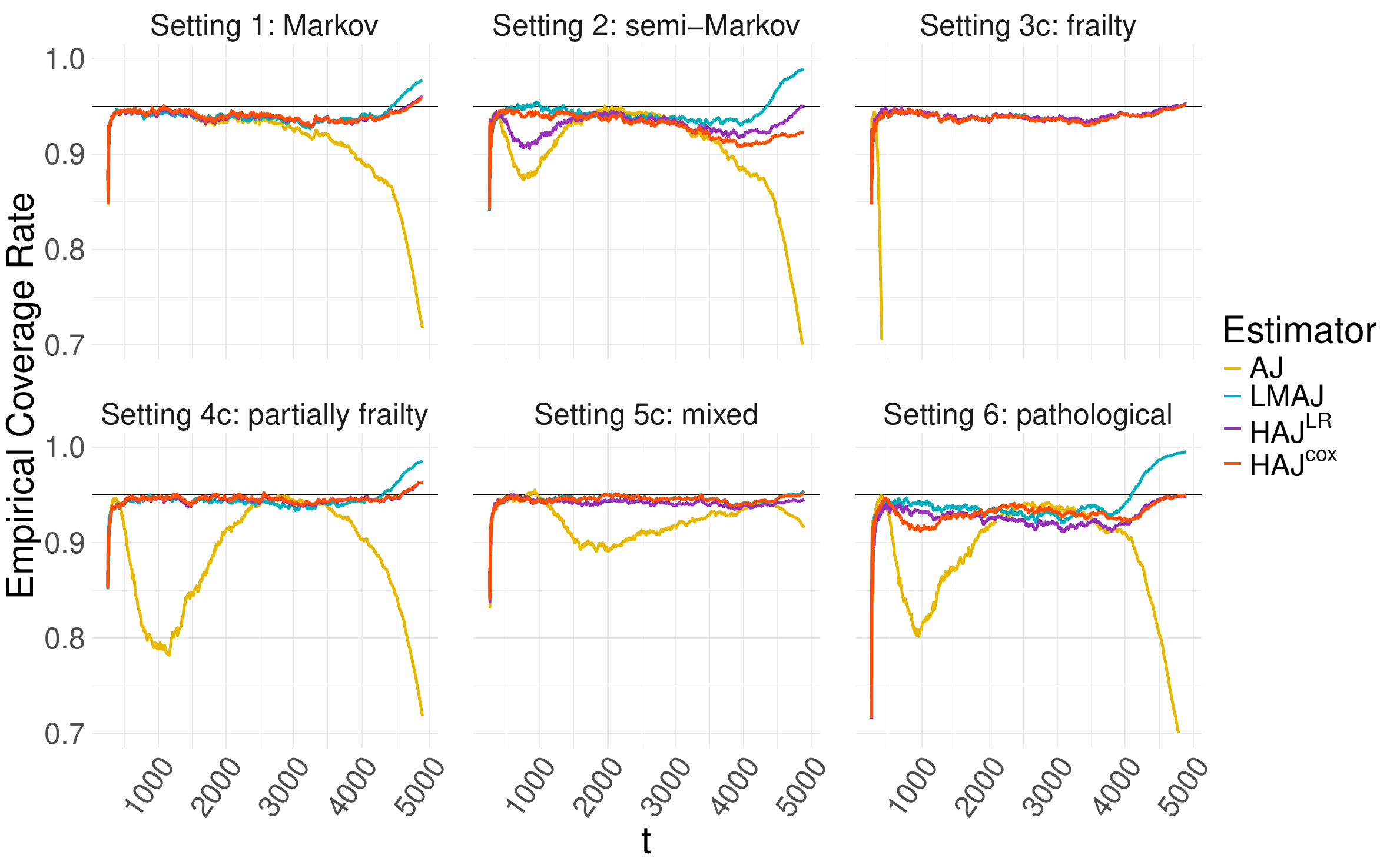}
\caption{Empirical coverage rate of the pointwise 95\% confidence intervals for the transition probability $P_{12}(245.57, t)$ for the process 1, 2, 3c, 4c, 5c and 6 for the AJ, LMAJ, HAJ$^{\text{LR}}$ and HAJ$^{\text{cox}}$ estimator. The black horizontal line is at 0.95.} 
\label{fig:empcov_12_205} 
\end{figure}

Lastly, the empirical coverage rate is considered as criterion.
It is depicted in Figure~\ref{fig:empcov_12_205} for $P_{12}(245.57, t)$. 
It is important to put the results for the later time points~$t$ into perspective.
Figures~D.3--D.14 in the supplement present the average number of subjects occupying each state over time across the 10000 simulation runs for the different settings. 
As expected, the number of subjects is increasing for state 3, while it is decreasing for the other two states with an exception for early time points in state~1. There are fewer than 50 subjects in state 1 and 2 after time point 2500 due to censoring and transitions to state $3$. 
With fewer people in the non-absorbing states $1$ and $2$, the variance of the estimator increases. 
The higher variance in later time points results in too conservative confidence intervals with empirical coverage rates exceeding 0.95 and even approaching 1. Another problem for the later time points is that the normal approximation might not hold true, yielding a lower empirical coverage rate in the case of the AJ estimator. Consequently, the interpretation of these later time points should be handled with caution.

Besides setting 1, the AJ estimator shows clear deviations from the nominal level.
The under-coverage can be explained by the bias.
In setting 3c, the AJ estimator even has an empirical coverage rate of 0 for most time points as seen in Figure~D.15 in the supplement. This further underlines the unsuitability of the AJ estimator for non-Markov processes.
Overall, the LMAJ and the HAJ estimators show an acceptable coverage in all settings closely to the 95\% level, with slight deviations for the HAJ$^{\text{LR}}$ estimator in setting~2 and all estimators in setting 6, which might be explained by the bias. This suggests that these three estimators provide reliable confidence intervals for most time points.

Altogether, the analysis revealed no definitive best estimator for all settings. The choice depends on the setting, the specific criterion and time point $t$ under consideration. Notably, the AJ estimator showed efficiency primarily in the Markov setting. In all other cases it exhibited a large bias and low values for the pointwise empirical coverage rate, indicating that its application should be limited to Markov processes. On the other hand, the LMAJ estimator demonstrated an overall good performance with favourable results in terms of bias and empirical coverage rate across all settings, but showed larger values for the variance in settings 1, 2, and 4c. 
The performance of the HAJ estimators seems to be rather similar in most settings.
Notable differences only occurred in the semi-Markov and the pathological setting, resulting in different conclusions:
The HAJ$^{\text{cox}}$ estimator demonstrated promising results with lower bias compared to the HAJ$^{\text{LR}}$ estimator in the semi-Markov setting. However, the HAJ$^{\text{LR}}$ estimator has also a higher variance, which leads to a similar RMSE compared to the AJ and HAJ$^{\text{LR}}$ estimators. On the other hand, the HAJ$^{\text{LR}}$ estimator displays a slightly smaller bias compared to the HAJ$^{\text{cox}}$ and AJ estimator together with a comparable variance in the pathological setting.
These results closely align with the relative power of the Cox-based and log-rank-based tests found in \cite{titman2022general} for these different scenarios, with the more powerful test performing better in each case.

Given the uncertainty of whether Markov assumptions hold in real-life scenarios, the HAJ estimators appear to be a favourable choice based on the results of this simulation study. Since the AJ estimator performs poorly in non-Markov settings, whereas the HAJ estimators perform well, and they have comparable values in the Markov and semi-Markov setting, there is little downside to selecting one of the HAJ estimators over the AJ estimator.
Furthermore, as the HAJ estimators outperform the LMAJ estimator in the Markov and semi-Markov setting, and demonstrate similar results in the non-Markov cases, they are an overall safe option. The only drawback is a slightly increased bias in some of the settings.

\subsection{Further Investigations}
In this section, we summarise further findings from the simulation study. 
This includes a comparison of the two tests for the Markov property, the influence of the degree of non-Markov behaviour, a comparison of different transitions, and the influence of the starting time.
More details on this matter can be found in Section~C in the supplement.

\paragraph{Comparison of Tests for the Markov Property}
The difference between the HAJ$^\text{LR}$ and HAJ$^\text{cox}$ estimator lies in the group of subjects included in the estimation, resulting from different test decisions.
In order to investigate the different performance of the two estimators in more detail, we analysed the tests in Section~C.1 in the supplement.
Here, we observe that the conclusions drawn from comparing the test results do not directly correspond to the results of the estimation.
In fact, the Cox method generally yields smaller p-values in most of the non-Markov cases than the log-rank-based test.
However, the resulting reduced bias of the estimator based on the Cox method is too small in most of the settings.

\paragraph{Influence of the Degree of non-Markov Behaviour}
The analysis of the influence of the degree of non-Markov behaviour in Section~C.2 in the supplement showed that the HAJ estimators work well for different degrees and forms of non-Markov behaviour. If there is only a slight deviation from non-Markov behaviour (settings a) or a transition is less travelled (transition $1 \rightarrow 3$ in setting 4c) the AJ estimator still works well, due to its advantage of a lower variance. 
 This should also be kept in mind when there are fewer subjects overall.

\paragraph{Comparison of Different Transitions}
Section~C.3 in the supplement illustrates the differences between the transitions and their complex interrelationships for the probability estimation within a model. The performance of the tests and estimators can vary across transitions within the same model. 
Additionally, the form of dependency on the history in one transition may influence the estimation of another transition in the same model, visible through the partially frailty model. 
The combination of those aspects makes it impossible to know the exact net effect of factors like the parameters of the transition intensity or dependency on the history. But it clearly shows that the model needs to be considered with all transitions as a whole for an overall decision on the best performing estimator. 

\paragraph{Influence of the Starting Time}
Section~C.4 in the supplement demonstrated that varying starting time points can affect the transition probability estimation, often altering only the distance between performance measure values of the estimators. However, in certain settings like the mixed process, also the ordering was changed, with the AJ estimator performing better initially but deteriorating as starting time increases. 
The tests reveal a slight decrease in detecting non-Markov transitions with increasing starting time.
Some of the results can be attributed to the sample size, since it is directly connected to the starting time.

\section{Conclusions and Outlook}
\label{sec:conl}


%

\vspace{0.3cm}

The overall aim of this paper was to explore several estimators for the transition probabilities in multi-state models across diverse settings, especially including various non-Markov scenarios. This exploration was conducted through a comprehensive simulation study using an illness-death model with recovery.
Besides the Aalen-Johansen estimator, typically used for Markov settings, the landmark Aalen-Johansen and the hybrid Aalen-Johansen estimator using a log-rank-based test were considered to accommodate non-Markov behaviour. Furthermore, a new version of the hybrid estimator applying a Cox model was introduced. 
Besides the Markov and semi-Markov setting, four distinct non-Markov settings were included to assess the estimators behaviour under different characteristics. 

%
In the simulation study, the Aalen-Johansen estimator performs well overall in the Markov setting. However, if that property is not fulfilled, the estimation is biased. 
It is important to note that this property must hold for the entire model, otherwise, bias can persist even if the specific transition is indeed Markov. 
In contrast, the landmark Aalen-Johansen estimator shows no bias and maintains an acceptable empirical coverage rate across all settings, albeit with higher variance due to data reduction compared to the standard Aalen-Johansen estimator. Especially in the pathological non-Markov model, it performed well.
The hybrid Aalen-Johansen estimators serve as a compromise between the two approaches, yielding favourable results across various measures and settings. While there are instances where the bias is higher, the RMSE and variance exhibit similar or smaller values compared to the landmark Aalen-Johansen estimator.
Particularly in the context of the partially non-Markov model, a hybrid estimator is preferred due to its ability to accommodate both Markov and non-Markov transitions effectively. 
Additionally, it performs similarly well in the Markov and semi-Markov case as the Aalen-Johansen estimator.
Overall, this makes the hybrid Aalen-Johansen estimators the most versatile among the ones considered, as they are well-suited for a wide range of settings.

Comparing the newly added version of the hybrid estimator incorporating a Cox method instead of the log-rank-based test for checking the Markov assumption shows that both estimators perform rather similar.
Further examination of the test results revealed that the Cox method frequently outperforms the log-rank-based test in capturing non-Markov transitions across most settings. Surprisingly, this superiority in testing does not necessarily translate into an improved estimation.

Overall, the simulation demonstrated the inherent complexity in analysing potentially non-Markov multi-state models. 
While the aforementioned conclusions regarding the estimators hold true in general, the choice between the original, landmark, and both hybrid Aalen-Johansen estimators depends on multiple factors. Some of these factors influence the results to the extent that the best-performing estimator may vary.
Among these influential factors are the form and degree of non-Markov behaviour, the different transitions, and the starting time.
Given that certain crucial information, such as adherence to the Markov assumption, may not be available in real-life data analysis, the simulation study suggests that both hybrid Aalen-Johansen estimators emerge as a good choice. 

Many of the influential factors ultimately revolve around the sample size, which likewise is the primary challenge of the landmark Aalen-Johansen estimator. In future research, it may be beneficial to expand the simulation study to include various sample sizes. 

Another aspect for consideration in future analyses concerns the hybrid estimator. While discussing the performance of the Aalen-Johansen estimator, an interesting observation is that this estimator can still yield reasonable results in non-Markov settings if the deviation from Markov behaviour is not too pronounced. Furthermore, a more precise test does not necessarily lead to a better estimator due to the importance of sample size. The testing thus far has been conducted at a significance level of 5\% without adjusting the p-values. 
In future analyses, the significance level could be changed. 

Reflecting on the outcomes of the hybrid Aalen-Johansen estimator employing the log-rank-based test within the pathological model, it becomes evident that the grid-test failed to detect non-Markov behaviour across several transitions. This limitation may stem from the grid's inadequate coverage of pivotal time points. 
An alteration involving the inclusion of relevant time points in the grid could be beneficial, potentially enhancing the testing procedures. Similarly, the starting time could be adjusted. 

Furthermore, some technical adjustments can be made in future analyses.
One of the adjustments involves the variance estimation and, consequently, the empirical coverage rate.
\citet{maltzahn2021hybrid} discuss the applicability of the Greenwood estimator in both Markov and non-Markov states. Despite \citet[p.~366]{glidden2002robust} stating that the variance of the transition matrix is the same for Markov and non-Markov data, the aforementioned authors suggest that in non-Markov states, the Greenwood estimator may not be applicable. Instead, they propose using a bootstrap approach. Since their comparison revealed only minor deviations between the bootstrap method and Greenwood estimator, the latter 
method was used in the study to streamline the computation time.
However, the bootstrap method could be preferred in future studies or real-world scenarios. 

Another aspect of the empirical coverage rate that can be improved in future analyses is the form of the confidence intervals. The utilisation of a normal approximation may not be justifiable, particularly for later time points where sample sizes are small. Considering this limitation, it would be beneficial to explore alternative methods for constructing confidence intervals. Options such as bootstrapping or using distributions better suited might offer more reliable intervals.

Lastly, another point of future research could be the comparison of the recently proposed kernel estimator of \cite{bladt2025conditional}, which also includes the incorporation of further covariates, to the hybrid Aalen-Johansen estimators computed for each of a set of discrete covariates.

\section*{Funding}
 \noindent M.\ Munko gratefully acknowledges funding by the German Research Foundation (Deutsche Forschungsgemeinschaft, DFG) - 314838170, GRK 2297 MathCoRe.

\section*{Acknowledgements}
    \noindent  D.\ Dobler would also like to thank his former affiliation, Department of Statistics, TU Dortmund University, and Research Center Trustworthy Data Science and Security, University Alliance Ruhr, Germany.

\appendix

\bibliographystyle{apalike} 
\bibliography{010_main.bbl}



\section*{Supplementary material}

# Supplementary Materials to Simulation Study to Compare Estimators for Multi-State Models in Potentially Non-Markov Processes

Carolin Drenda[a], Dennis Dobler[b], Merle Munko[c], Andrew Titman[d]

[a]*TU Dortmund University, August-Schmidt-Straße 1, Dortmund, 44227, Germany*
[b]*RWTH Aachen University, Templergraben 55, Aachen, 52026, Germany*
[c]*Otto von Guericke University, Universitätsplatz 2, Magdeburg, 39106, Germany*
[d]*Lancaster University, LA1 4YW, Lancaster, United Kingdom*

## A. Details on the Log-Rank-Based Test

If not stated otherwise, Titman and Putter (2020a) is used as a source for this section.

### A.1. Local Test

The local test focusses on a specific time $s \in \mathcal{T}$ and state $l \in \mathcal{S}$, yielding two groups of subjects $\mathcal{L} = \{i : \tilde{X}_i(s) = l, \tilde{Y}_i(s) = 1\}$ and $\mathcal{L}^C = \{i : \tilde{X}_i(s) \neq l, \tilde{Y}_i(s) = 1\}$. Under the Markov assumption, the transition intensities would be the same in both groups for all $t > s$. This leads to the null hypothesis

$$H_0 : \alpha_{hj}\big(t|\tilde{X}(s) = l\big) = \alpha_{hj}\big(t|\tilde{X}(s) \neq l\big),\ t > s,\ t \in \mathcal{T},$$

versus a general alternative. Let $\delta^i_l(s) = \mathbb{1}\{\tilde{X}_i(s) = l\}$ represent the indicator function for the $i$th subject being in the state of interest $l$ at time $s$. The *log-rank statistic* for each transition $h \to j$ is defined as

$$U_s^{(l)}(h,j) := \sum_{i=1}^{n} \int_s^{\tau} \left\{ \delta^i_l(s) - \frac{\sum_{m=1}^n \delta^m_l(s)\tilde{Y}^m_h(u)}{\sum_{m=1}^n \tilde{Y}^m_h(u)} \right\} d\tilde{N}^i_{hj}(u). \qquad (A.1)$$

Let $\mathcal{R}_l$ be the set of states reachable from state $l$, and $\mathcal{R}_{l^c} = \bigcup_{l' \neq l} \mathcal{R}_{l'}$, the set of states reachable from all other states $l' \neq l$. The test statistic is

defined for transitions from state $h \in \mathcal{R}_l \bigcap \mathcal{R}_{l^c}$, signifying transitions from state $h$ can be reached from state $l$ and at least one other state $l' \neq l$. This is necessary since, otherwise, the test statistic would be uniformly 0. Standardising the statistic to

$$\bar{U}_s^{(l)}(h,j) := \frac{U_s^{(l)}(h,j)}{\sqrt{\widehat{\text{Var}}\big(U_s^{(l)}(h,j)\big)}} \tag{A.2}$$

enables the comparison to a standard normal distribution for each individual hypothesis. The derivation of the estimated variance is provided in A.5.

*A.2. Global Test*

In the previous section only a fixed time point $s$ was considered. To develop a global version for a set state $l$, a range of time points $[t_0, t_{\max}] \subset [0, \tau]$ is examined. The null hypothesis extends to

$$H_0 : \alpha_{hj}\big(t|\tilde{X}(s) = l\big) = \alpha_{hj}\big(t|\tilde{X}(s) \neq l\big) \ \forall \ s \in [t_0, t_{\max}], \ s \leq t \leq \tau.$$

The process $\big(\bar{U}_s^{(l)}(h,j), \ s \in [t_0, t_{\max}]\big)$ converges to a zero mean Gaussian process with consistently estimable covariance function. Details regarding the form and estimation of this covariance function are also provided in A.5. To receive a global test statistic, the supremum of the process $\sup_{s \in [t_0, t_{\max}]} |\bar{U}_s^{(l)}(h,j)|$ is used.

For the actual calculations, there is an important consideration. The value of $\bar{U}_s^{(l)}(h,j)$ may change with any alteration in the group of subjects occupying state $h$. Such changes could stem from subjects leaving or entering the state, or from a subject in this state being censored. Consequently, determining $\bar{U}_s^{(l)}(h,j)$ at each time point is computationally expensive and it is preferred to use a time grid $s_1, ..., s_G$ for the evaluation. Subsequently, a summary test statistic over the whole time grid can be calculated as $\max_{s_g} |\bar{U}_{s_g}^{(l)}(h,j)|$.

Another problem associated with the global test arises when there are only a few patients either in or not in the qualifying state. This situation can for example occur at the beginning or the end of a study, potentially resulting in an undefined or unstable log-rank statistic. To avoid these problems, the time range $[t_0, t_{\max}]$ can be truncated with the help of the weights

$$\omega_l(s) := \frac{\sqrt{\gamma(s) n_l^{(1)}(s) n_l^{(0)}(s)}}{n_l^{(1)}(s) + n_l^{(0)}(s)}, \tag{A.3}$$

with

$$n_l^{(1)}(s) := \sum_{i=1}^{n} \delta_l^i(s) H_i(s) \tag{A.4}$$

$$n_l^{(0)}(s) := \sum_{i=1}^{n} \bigl(1 - \delta_l^i(s)\bigr) H_i(s) \tag{A.5}$$

$$\gamma(s) := \sum_{h,j} \tilde{N}_{hj}(\tau) - \tilde{N}_{hj}(s) \tag{A.6}$$

where Equation (A.4) is the number of uncensored subjects that are in state $l$ at time $s$, Equation (A.5) the number of uncensored subjects that are not in state $l$ at time $s$, and Equation (A.6) is the total number of transitions after time $s$. As suggested by the authors, only time points with $\omega_l(s) > 2.5$ are used to ensure that the estimated power to detect non-Markov behaviour is sufficient. For more details about the weights refer to S6 in Titman and Putter (2020b).

### A.3. Transition Specific Statistic

In a multi-state model, there can exist more than two qualifying states $l$ that are relevant to a transition $h \to j$. For example in the illness-death model with recovery, considering the transition $1 \to 2$, the qualifying state can be either 1 or 2. Then, $\mathcal{R}_j^*$ represents the set of states from which state $j$ is reachable. Starting again with a local test for a single time point $s$, the score statistic $U_s^{(l)}(h,j)$ is defined as in Equation (A.1) for each $l \in \mathcal{R}_j^*$. These test statistics for the different qualifying states $l$ are aggregated into a vector $\boldsymbol{U}_s$, where the notation for the transition $h \to j$ is omitted for improved readability from this point forward. The covariance between the entries of this vector correspond to the entries of the matrix $\boldsymbol{\Psi}_s$ with $(ll')th$ entry given by

$$\sum_{i=1}^{n} \int_{(s,\tau]} \left\{ \delta_l^i(s) - \frac{\sum_{m=1}^{n} \delta_l^m(s) \tilde{Y}_{hj}^m(t)}{\sum_{m=1}^{n} \tilde{Y}_{hj}^m(t)} \right\} \left\{ \delta_{l'}^i(s) - \frac{\sum_{m=1}^{n} \delta_{l'}^m(s) \tilde{Y}_{hj}^m(t)}{\sum_{m=1}^{n} \tilde{Y}_{hj}^m(t)} \right\} d\hat{\boldsymbol{A}}(t), \tag{A.7}$$

where $\hat{\boldsymbol{A}}(t)$ is the Nelson-Aalen estimator for the cumulative transition intensity (Titman and Putter, 2020b). Asymptotically, under the null hypothesis,

3

$\boldsymbol{U}_s$ is centred. Therefore, the asymptotic distribution of the test statistic

$$\boldsymbol{U}_{s(r)}^\intercal \boldsymbol{\Psi}_{s(r,r)}^{-1} \boldsymbol{U}_{s(r)} \tag{A.8}$$

is $\chi^2_{r^*-1}$, with $r^* = |\mathcal{R}_j^*|$. Here, $\boldsymbol{U}_{s(r)}$ is the vector $\boldsymbol{U}_s$ with the $r$th element removed and $\boldsymbol{\Psi}_{s(r,r)}$ with removed $r$th row and column from $\boldsymbol{\Psi}_s$. The choice of $r$ is arbitrary within $\mathcal{R}_j^*$. This removal is necessary to prevent the covariance matrix from being singular and therefore not invertible.

Expanding the test statistic to cover multiple time points $s$ for a global test can be achieved by summarising (A.8) as in A.2, as follows:

$$\max_{g \in \{1,...,G\}} \boldsymbol{U}_{s_g(r)}^\intercal \boldsymbol{\Psi}_{s_g(r,r)}^{-1} \boldsymbol{U}_{s_g(r)}. \tag{A.9}$$

It tests the hypothesis

$$H_0 : \alpha_{hj}(t|\tilde{X}(s) = l) = \alpha_{hj}(t|\tilde{X}(s) \neq l) \ \forall \ s \in [t_0, t_{\max}], \ s \leq t \leq \tau, \ \forall \ l \in \mathcal{R}_j^*.$$

*A.4. Null Distribution*

There are different ways to determine the asymptotic null distribution for the test statistic (A.9). Here, a wild bootstrap is used, since it exhibits a good small sample approximation (Beyersmann et al., 2013). The realisations of (A.1) are

$$u_s^{(l)}(h,j) := \sum_{i=1}^n \sum_{g=\tilde{N}_{hj}^i(s)+1}^{\tilde{N}_{hj}^i(\tau)} \left( \delta_l^i(s) - \frac{\sum_{m=1}^n \delta_l^m(s)\tilde{Y}_h^m(t_{ig})}{\sum_{m=1}^n \tilde{Y}_h^m(t_{ig})} \right), \tag{A.10}$$

where $t_{ig}$ corresponds to the $g$th direct transition $h \to j$ for subject $i$. For the wild bootstrap version, (A.10) is multiplied with independent and identically distributed centred Poisson random variables $\mathcal{G}_{ig}$, $i = 1,...,n$, $g = \tilde{N}_{hj}^i(s) + 1,...,\tilde{N}_{hj}^i(\tau)$, with mean 0 and variance 1. This leads to the following form:

$$u_s^{(l)*}(h,j) := \sum_{i=1}^n \sum_{g=\tilde{N}_{hj}^i(s)+1}^{\tilde{N}_{hj}^i(\tau)} \left( \delta_l^i(s) - \frac{\sum_{m=1}^n \delta_l^m(s)\tilde{Y}_h^m(t_{ig})}{\sum_{m=1}^n \tilde{Y}_h^m(t_{ig})} \right) \mathcal{G}_{ig}.$$

The bootstrap procedure is iterated $n_B$ times for each time point $s \in \{s_1,...,s_G\}$ and one bootstrap test statistic is calculated per iteration. Subsequently, the p-value is determined by the proportion of bootstrap realisations that are the same as or more extreme than the observed test statistic (A.9). If the p-value is smaller than a specified $\alpha \in [0,1]$, non-Markov behaviour is assumed. Otherwise the Markov property is assumed.

*A.5. Covariance*

The asymptotic covariance between two test statistics of the form (A.1) for time points $s$ and $s'$, $s \leq s'$ is given by

$$\text{Cov}\Big(U_s^{(l)}(h,j), U_{s'}^{(l)}(h,j)\Big) :=$$

$$\sum_{i=1}^n \int_{s'}^{\tau} \tilde{Y}_h^i(t) \left\{ \delta_l^i(s) - \frac{\sum_{m=1}^n \delta_l^m(s)\tilde{Y}_h^m(t)}{\sum_{m=1}^n \tilde{Y}_h^m(t)} \right\} \left\{ \delta_l^i(s') - \frac{\sum_{m=1}^n \delta_l^m(s')\tilde{Y}_h^m(t)}{\sum_{m=1}^n \tilde{Y}_h^m(t)} \right\} d\hat{\mathbf{A}}(t)$$

(A.11)

where $\hat{\mathbf{A}}(t)$ is the Nelson-Aalen estimator for the cumulative transition intensity. To estimate (A.11) define $n_v := \sum_{i=1}^n \tilde{Y}_h^i\big(t_{hj}^{(v)}\big)$ as number of subjects at risk for each unique transition time $t_{hj}^{(v)} \geq s$. This number can also be expressed as

$$n_v = n_{v11} + n_{v01} + n_{v10} + n_{v00}$$

with

$$n_{v11} := \sum_{i=1}^n \tilde{Y}_{hj}^i\big(t_{hj}^{(v)}\big) \delta_l^i(s) \delta_l^i(s'),$$

$$n_{v01} := \sum_{i=1}^n \tilde{Y}_{hj}^i\big(t_{hj}^{(v)}\big) \big(1 - \delta_l^i(s)\big) \delta_l^i(s'),$$

$$n_{v10} := \sum_{i=1}^n \tilde{Y}_{hj}^i\big(t_{hj}^{(v)}\big) \delta_l^i(s) \big(1 - \delta_l^i(s')\big),$$

$$n_{v10} := \sum_{i=1}^n \tilde{Y}_{hj}^i\big(t_{hj}^{(v)}\big) \big(1 - \delta_l^i(s)\big) \big(1 - \delta_l^i(s')\big).$$

Then, the estimate for a given dataset is denoted by

$$\widehat{\text{Cov}}\Big(U_s^{(l)}(h,j), U_{s'}^{(l)}(h,j)\Big) := \sum_{v: t_{hj}^{(v)} \geq s'} \frac{n_{v11} n_{v00} - n_{v01} n_{v10}}{n_v^2}.$$

To estimate the variance for (A.2) set $s' = s$, which yields

$$\widehat{\text{Var}}\Big(U_s^{(l)}(h,j)\Big) := \sum_{v: t_{hj}^{(v)} \geq s} \frac{n_{v11}(n_v - n_{v11})}{n_v^2}.$$

The process $\left(\bar{U}_s^{(l)}, s \in [t_0, t_{\max}]\right)$ converges under the null hypothesis to a zero mean Gaussian process with covariance function

$$\frac{\operatorname{Cov}\left(U_s^{(l)}(h,j), U_{s'}^{(l)}(h,j)\right)}{\sqrt{\operatorname{Var}\left(U_s^{(l)}(h,j)\right)\operatorname{Var}\left(U_{s'}^{(l)}(h,j)\right)}}$$

(Titman and Putter, 2020a).

## B. Details on the Markov Model

In this section, we provide the calculations that lead to the modelling of the transition times in Section 4.2.1. Therefore, the transition subscript $hj$ is omitted during the derivation to improve readability. The conditional survival probabilities can be calculated as

$$\mathbb{P}(T > t | T > t_0) = \exp\bigl(-A(t) + A(t_0)\bigr), \tag{B.1}$$

where $A$ denotes the cumulative hazard function. Conditioning on $T > t_0$ is necessary, as the subject has not been censored in the preceding step; otherwise, the algorithm would have already terminated. Using the transition intensities of the Weibull distribution with parameters $a, b$ for Equation (B.1) yields

$$\mathbb{P}(T > t | T > t_0) = \exp\left(-at^b + at_0^b\right).$$

Hence, for $u$ drawn from the uniform distribution $U \sim \mathcal{U}[0, 1]$, we obtain

$$u = \exp\left(-at^b + at_0^b\right) \quad \Leftrightarrow \quad t = \left(t_0^b - \frac{1}{a}\log(u)\right)^{\frac{1}{b}}.$$

## C. Details on the Simulation Results

*C.1. Comparison of Tests for the Markov Property*

In the following, the test decisions of the log-rank-based test and the Cox model are compared to get further insights on the two testing procedures. In Figure C.1, the empirical distribution of the p-values over the 10000 simulation replicates is shown for the test of the transition $1 \to 2$. A p-value smaller than 0.05 suggests that the transition is non-Markov. For the Markov setting as well as setting 4c, the p-values are uniformly distributed, indicating that the null hypothesis is rejected in about 5% of the cases as expected for those settings with the applied significance level. In setting 4c, only transition $2 \to 1$ is modelled as non-Markov. The decision in favour of the Markov property is therefore correct. For settings 2, 3c and 5c, the Cox method generally yields smaller p-values than the log-rank-based test, thus detecting deviations from Markov behaviour more frequently. The log-rank-based test has more power to detect the effect only in setting 6. The results concerning the power in the different settings are consistent with the findings reported by Titman and Putter (2020a) and support the idea of improving the HAJ estimator by using the Cox method.

The higher power of the test manifests as lower biases for the estimators, as seen in Figure 5. However, the difference in bias between the two HAJ estimators across different settings, while present, is relatively small compared to the variance value itself. In settings 3c and 5c, where the variance values are comparable (Figure 6), this similarity in bias leads to nearly identical RMSE values (Figure 7), making it challenging to make a clear decision when considering only this measure as criterion.

In setting 2, the advantage of a lower bias for the HAJ$^{\text{cox}}$ estimator is compensated by a higher variance, resulting in a similar RMSE. Therefore, the test results can be seen as a bias-variance trade-off as well, with a more accurate test decision leading to a lower bias but potentially higher variance due to the data reduction. This implies that improving the test decision does not necessarily improve the overall performance of the estimator, particularly if the reduction in bias compensates for the increase in variance.

Another aspect to consider is that, based on the test decision, the landmarking is performed. However, the Cox model utilises the time of the most recent entry into the current state as an aspect of non-Markov behaviour. In contrast, the applied landmarking method for the final estimation focusses on being in a state at a given time. It is possible that the time of entry is a

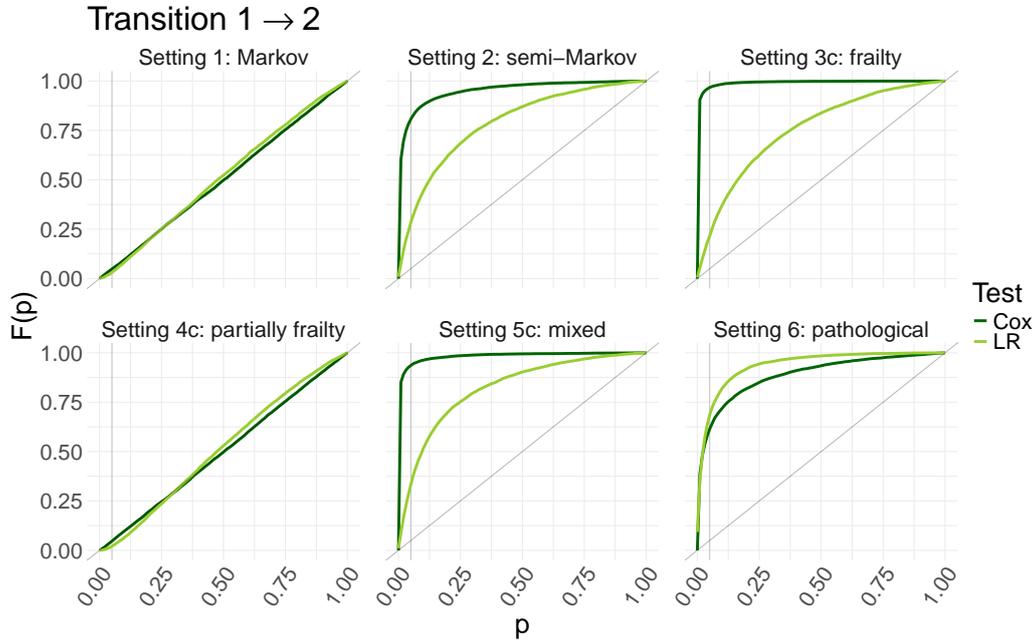

Figure C.1: Empirical distribution of the p-values for the log-rank-based test (LR) and the Cox method (Cox) over 10000 simulation runs for transition $1 \to 2$ for settings 1, 2, 3c, 4c, 5c and 6. The grey, vertical line is at 0.05.

better way to describe the non-Markov behaviour in some cases, which ultimately is not used to model the transition probability. Thus, the advantage of the more powerful test is only partially utilised.

But, there exists also a drawback in the applied log-rank version. For the testing, the maximum of test statistics over a grid of time points is used. It might be another time point than the landmark time point, that actually leads to the rejection of the null hypothesis and therefore, another time point that captures a group of subjects detecting the non-Markov behaviour. While the time point that led to the rejection could also be used as landmark time, its relevance depends on the research question and therefore is not always of interest.

Overall, the testing aspect of the hybrid estimators merely serves as a tool and is not utilised in a one-to-one manner during the subsequent estimation process. Hence, the conclusions drawn from comparing the test results do not directly correspond to the results of the estimation. This also explains why the new estimator does not really show an improvement, despite showing

favourable test results in most settings.

*C.2. Influence of the Degree of non-Markov Behaviour*

In this section, the results for different variances and, consequently, the degrees of non-Markov behaviour are compared. The empirical distributions of the p-values for both testing methods for settings 5a-c for the transition $1 \to 2$ are depicted in Figure C.2. As a reminder, the variances for the frailty variable are in ascending order with values 0.5, 1 and 2.

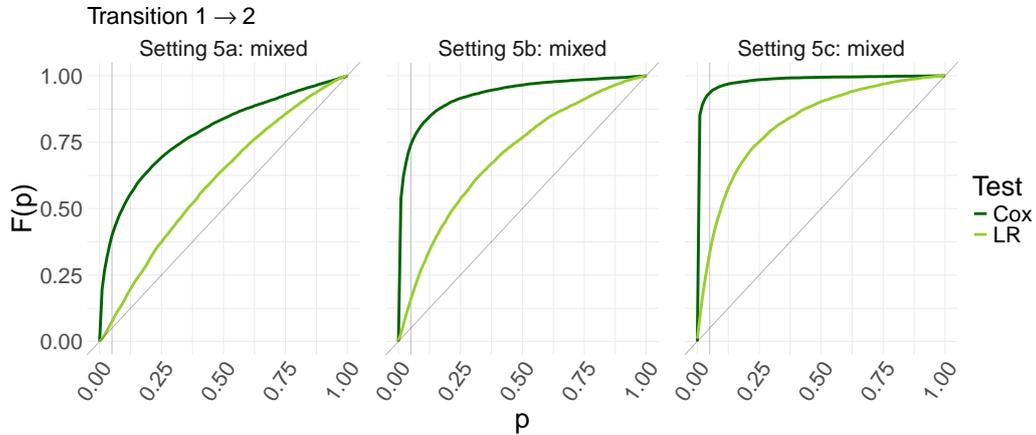

Figure C.2: Empirical distribution of the p-values for the log-rank-based test (LR) and the Cox method (Cox) over 10000 simulations for transition $1 \to 2$ for settings 5a-c. The grey vertical line is at 0.05.

As assumed, with increasing variance for the Gamma frailties and therefore increasing perturbation of the transition intensities, the rejection rates are higher for both tests. Across all considered settings, the Cox method detects the non-Markov transition more often than the log-rank-based test, consistent with the findings in the previous section. Similar outcomes are observed for settings 3a-c seen in Figure C.3.

As in the previous section, the better test results generally lead to a smaller bias, depicted in Figure C.4 for settings 5a-c.

The RMSE for settings 5a-c for transition $1 \to 2$ in Figure C.5 illustrates, that, if the deviation from Markov behaviour is large enough, the AJ estimator is outperformed by the other estimators. In setting 5a, where the frailty variance is the lowest, the HAJ$^{\text{LR}}$ and AJ estimator demonstrate a comparable RMSE. However, the bias for the AJ estimator is higher than

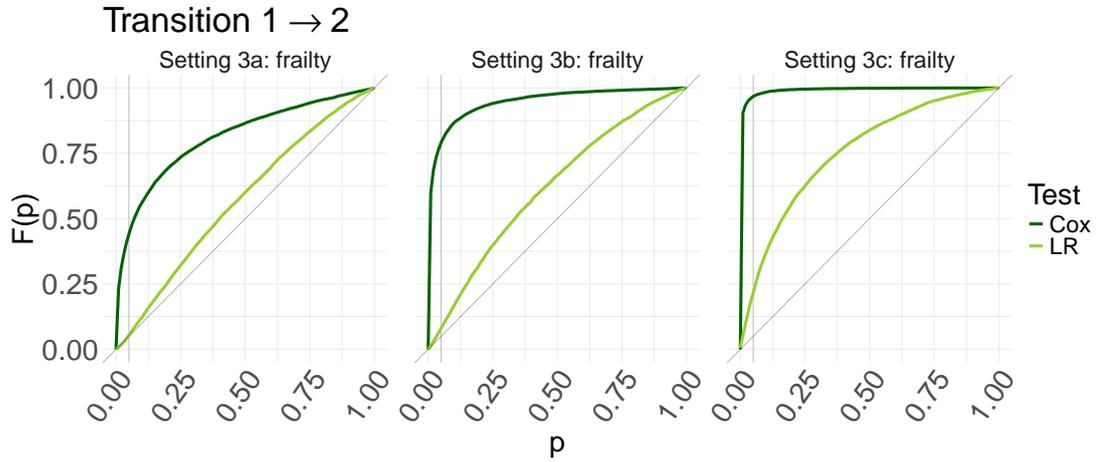

Figure C.3: Empirical distribution of the p-values for the log-rank-based test (LR) and the Cox method (Cox) over 10000 simulation runs for scenarios 3a-3c. The grey vertical line is at 0.05.

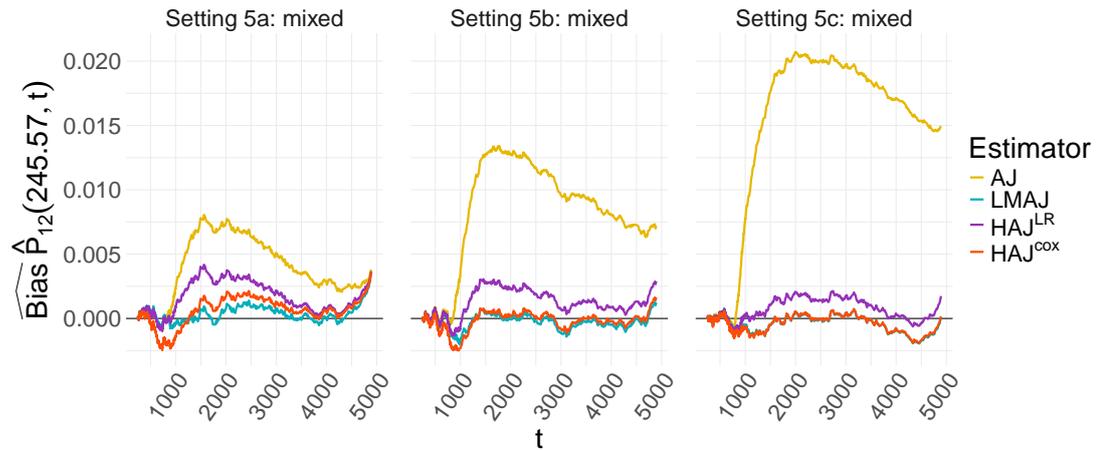

Figure C.4: Bias for $\hat{P}_{12}(s,t)$ with increasing frailty variance in settings 5a-c for the AJ, LMAJ, HAJ$^{\text{LR}}$ and HAJ$^{\text{cox}}$ estimator.

for the other estimators, as depicted in Figure C.4. With increasing frailty variance, the deviation from Markov behaviour is pronounced enough that the bias-variance trade-off is in favour of the LMAJ and the HAJ estimators.

When considering the empirical coverage rate as another criterion (Figure

11

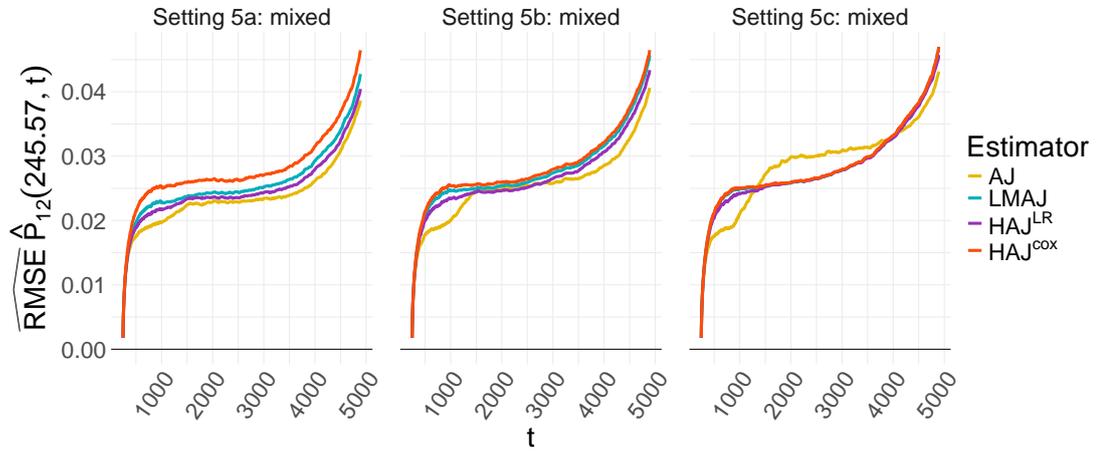

Figure C.5: RMSE for $\hat{P}_{12}(245.57, t)$ with increasing frailty variance for settings 5a-c: mixed for the AJ, LMAJ, HAJ$^{LR}$ and HAJ$^{cox}$ estimator.

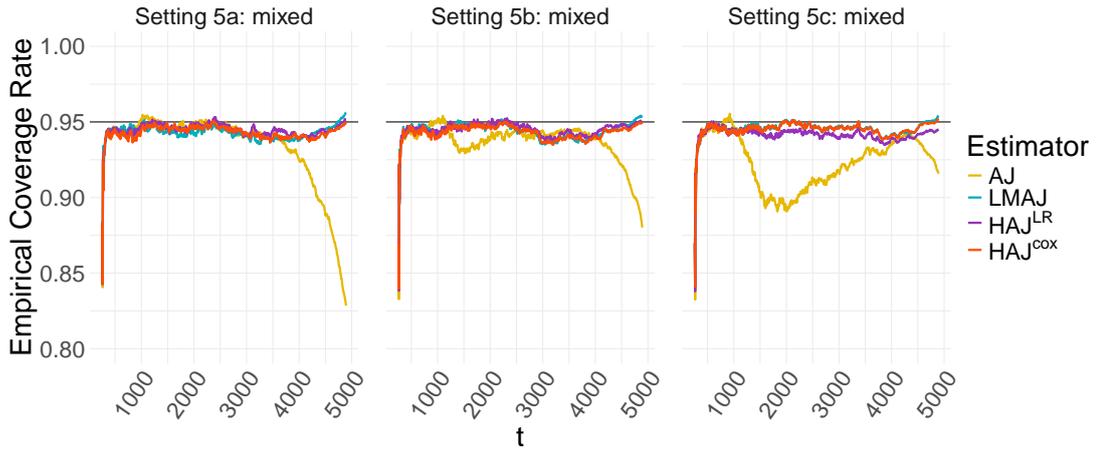

Figure C.6: Empirical coverage rate of the pointwise 95% confidence intervals for the transition probability $P_{12}(245.57, t)$ for settings 5a-c for the AJ, LMAJ, HAJ$^{LR}$ and HAJ$^{cox}$ estimator. The black horizontal line is at 0.95.

C.6), in setting 5a, the performance of the AJ estimator is similar to the other estimators and close to the 0.95 level for small and moderate time points. This suggests that the reliability of the confidence intervals aligns well with theoretical expectations. In settings 5b and c, however, the AJ estimator shows considerably lower coverage rates than the other three estimators.

12

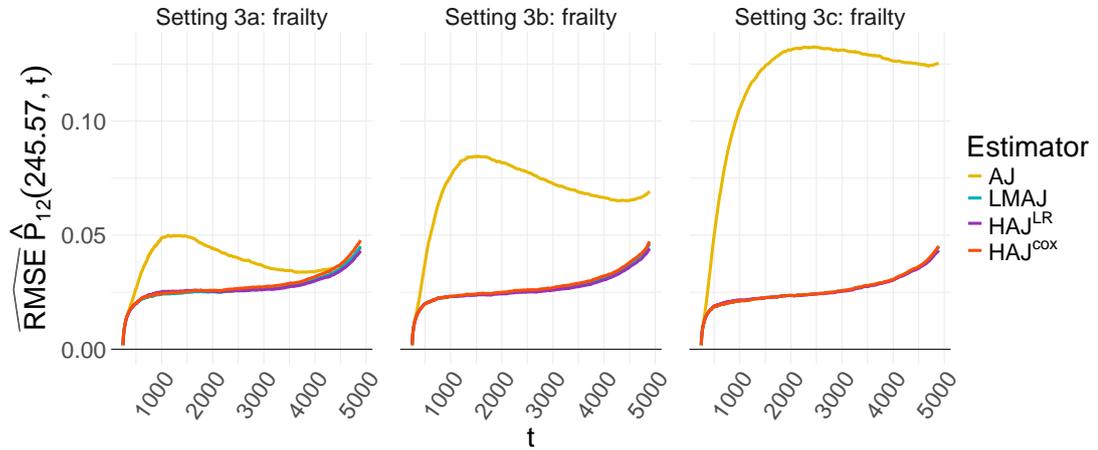

Figure C.7: RMSE for $\hat{P}_{12}(245.57, t)$ with increasing frailty variance for settings 3a-c for the AJ, LMAJ, HAJ$^{\text{LR}}$ and HAJ$^{\text{cox}}$ estimator.

Therefore, small deviations from Markov behaviour are tolerable for the AJ estimator with acceptable values in the RMSE and the empirical coverage rate. But with increasing degree, the AJ estimator is not recommended.

Another observation from Figure C.5 is that, with increasing non-Markov behaviour, the curves of the LMAJ and the HAJ estimators approach each other. The larger bias of the HAJ$^{\text{LR}}$ estimator and the larger variance of the LMAJ and HAJ$^{\text{cox}}$ estimators balance out, resulting in a similar RMSE. Therefore, the cases in which the HAJ$^{\text{LR}}$ estimator is favourable over the LMAJ estimator depend on the degree of non-Markov behaviour. When it is not too strong, a larger sample size has more influence on decreasing the RMSE than the correct test decision. However, this influence reduces with increasing degree.

For the frailty model in setting 3a-c, the RMSE values of the AJ estimator are already higher in the setting 3a than for the other estimators (Figure C.7). Likewise, the empirical coverage rate is unacceptably lower in the setting with lowest frailty variance (Figure C.8). In this setting, smaller perturbations than in setting 5 are sufficient that the LMAJ and HAJ estimators outperform the AJ estimator. This is plausible, since in setting 5, non-Markov behaviour is only introduced after time point 607.54, whereas in setting 3, the data is simulated as non-Markov for all time points.

Lastly, settings 4a-c are examined. The RMSE for $\hat{P}_{hj}(245.57, t)$ for all

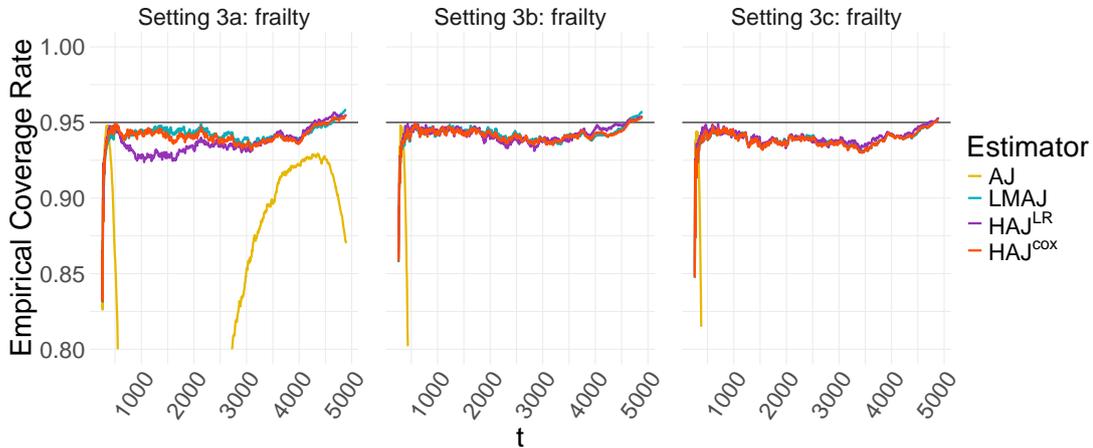

Figure C.8: Empirical coverage rate of the pointwise 95% confidence intervals for the transition probability $P_{12}(245.57, t)$ for settings 3a-c for the AJ, LMAJ, HAJ$^{\text{LR}}$ and HAJ$^{\text{cox}}$ estimator.

transitions is depicted in Figure C.9. The first row shows the results for the previously considered transition $1 \to 2$. Similar to setting 5, with increasing frailty variance, the efficiency of the HAJ estimators over the AJ estimator becomes more apparent. The same applies for transitions $1 \to 3$ and $2 \to 3$ in the second and last row. It should be noted, that these transitions were modelled according to the Markov assumption. When looking at the non-Markov transition $2 \to 1$ in the third row, the HAJ estimators already perform clearly better than the AJ and LMAJ estimator in setting b.

Nevertheless, another aspect becomes apparent in this context. The scale parameter for this transition is the lowest out of the four with $\hat{a}_{13} = 0.0003$. Due to the small parameter and the restriction that it can be at most observed once per subject, this is modelled as the least direct travelled transition. Care must be taken with the interpretation here, as a sample path $1 \to 2 \to 3$, for example, is also included in the estimation of this transition probability. But this indicates the importance of the sample size for less travelled transitions, since the AJ estimator yields similar RMSE values as the HAJ estimators in setting 4a) and 4b) due to a lower variance. This is also supported by the fact that the AJ estimator tends to have lower RMSE values for later time points. Since there are in general fewer observations later on, less reduction can improve the performance in comparison to the other estimators.

Overall, across all transitions and all frailty variances in the partially

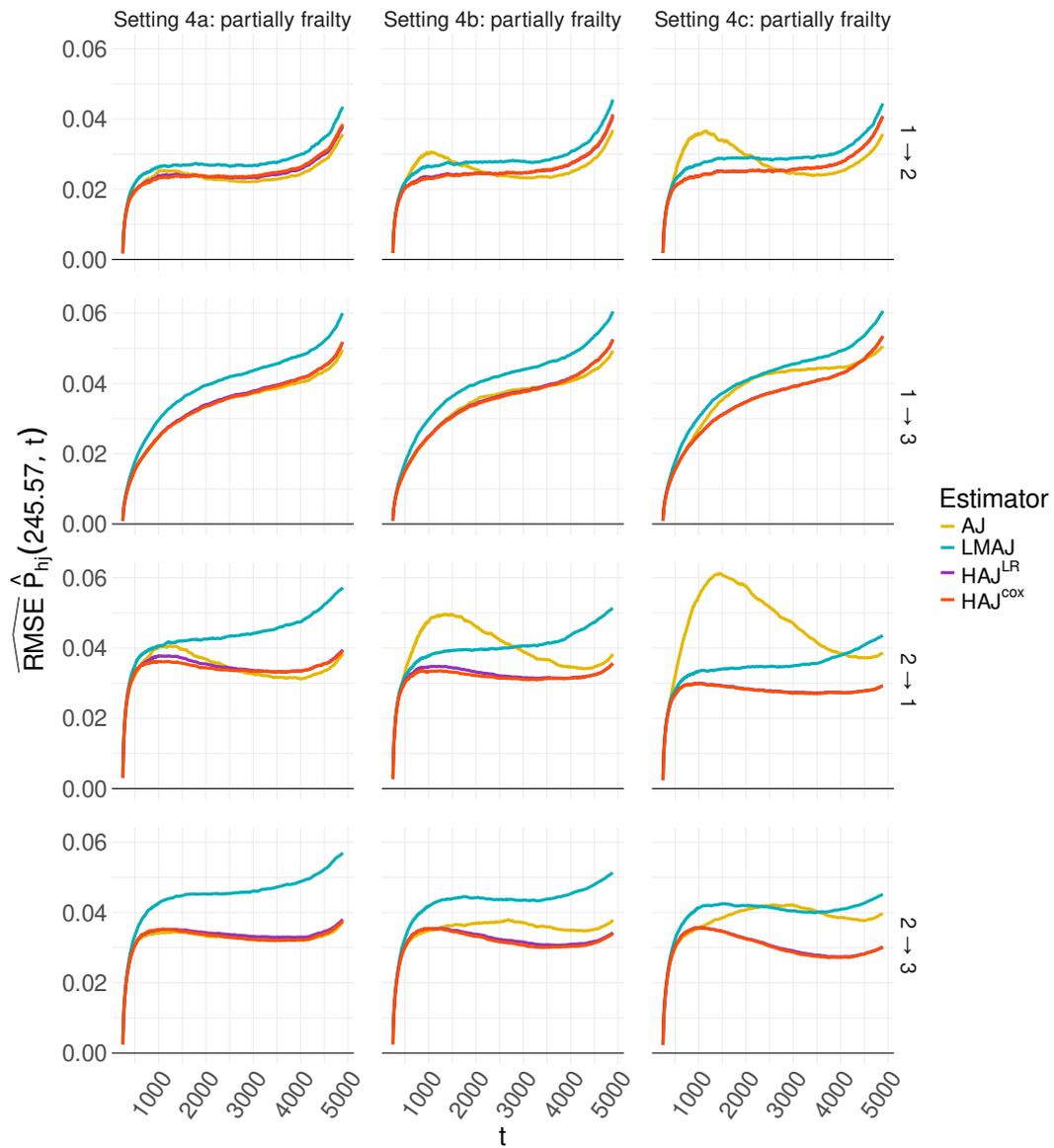

Figure C.9: RMSE for $\hat{P}_{hj}(245.57, t)$ for all transitions with increasing frailty variance for settings 4a-c: partially frailty, for the AJ, LMAJ, HAJ$^{\text{LR}}$ and HAJ$^{\text{cox}}$ estimator.

frailty model, the HAJ estimators yield a lower RMSE than the LMAJ. In this setting, the LMAJ estimator clearly is not advisable due to the unnecessary

15

reduction of subjects. The comparison to the AJ estimator reveals that the best-performing estimator, in terms of the RMSE, depends on the transition, degree of non-Markov behaviour, and time point $t$. But as mentioned, since the bias is lower, the HAJ estimators are preferred over the AJ estimator.

*C.3. Comparison of Different Transitions*

Starting with interesting differences in the test procedures, the empirical distribution of the p-values for the semi-Markov model for all transitions is depicted in Figure C.10. Especially noticeable is that for transition $2 \to 1$, the p-values for both tests are uniformly distributed indicating decisions in favour of the Markov behaviour according to the significance level. Taking a look at the parameter values for this transition explains this fact. The parameter $\hat{b}_{21}$ has a value of 1.0228. In the case of $\hat{b}_{21} = 1$, the time generation for the semi-Markov model in Equation (6) actually corresponds to that of the Markov model. The tests, therefore, do not have enough power to detect such a small difference. Conversely, for other transitions, the disparity is significant enough to lead to varying degrees of rejection of Markov behaviour. These rates may also be related to the size of parameter $\hat{b}_{hj}$, which ranges between 0.7453 and 0.8682 for these transitions. This indicates that the tests are sensitive to the different parameters of the transition intensities.

One important consideration when examining the various transitions within the same model is that the estimation of the transition probability relies on multiple transition intensities, rather than solely on the corresponding one. Therefore, it is possible, that the estimation of the transition intensity is unbiased, whereas the estimated probability shows some bias. An example for this is seen in Figure 5. Here, among others, the bias for transition $1 \to 2$ for setting 4c, the partially frailty model, is depicted. Although this transition is modelled as Markov (the frailty variable is only used for transition $2 \to 1$), the AJ estimator is biased. In a full Markov model, the bias would be much smaller as seen in the same figure. This implies, that it is important to consider the model as a whole. Additionally, it is necessary for the model to be completely Markov in order to prefer the AJ estimator over the other estimators.

In the case that the tests do not reject Markov behaviour for all transitions, both HAJ estimators are equal to the AJ estimator. If the null hypothesis is rejected for all transitions, they are in fact the LMAJ estimator. Therefore, test decisions for the combination of the four transitions together are of interest. The frequency of the most common combinations

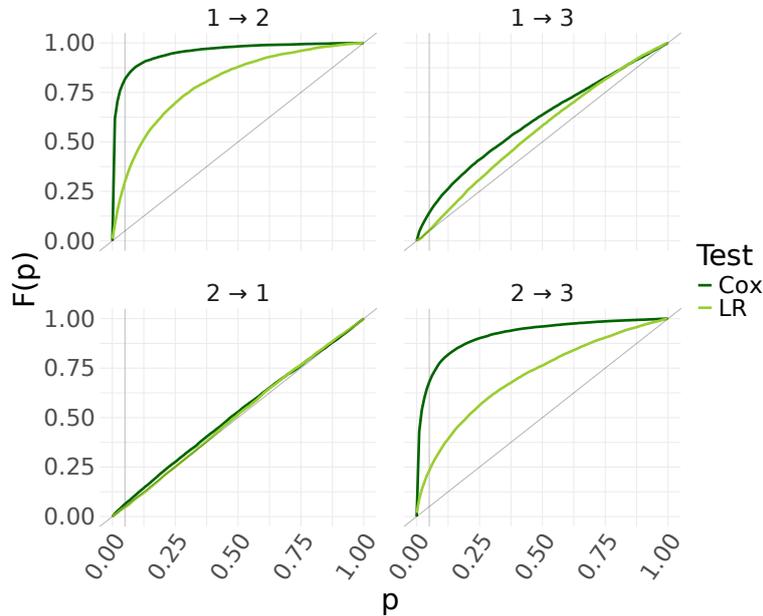

Figure C.10: Empirical distribution of the p-values for the log-rank-based test (LR) and the Cox method (Cox) over 10000 simulations over 10000 simulation runs for setting 2: semi-Markov for all transitions. The grey vertical line is at 0.05.

for model 1, 2, 3c, 4c, 5c and 6 are displayed in Figure C.11. The transitions are ordered as $1 \to 2$, $1 \to 3$, $2 \to 1$ and $2 \to 3$. A p-value smaller than 0.05 is encoded as 1, otherwise, it is encoded as 0. The combination 0011 for example means that transitions $1 \to 2$ and $1 \to 3$ are considered to be Markov and all subjects are used, whereas for transitions $2 \to 1$ and $2 \to 3$, landmarking is applied. Since different settings yield distinct most frequent combinations, the displayed combinations vary depending on the setting.

As expected, the combination 0000 is the most common one for setting 1 for both methods. Since the p-values are not adjusted and the tests have a significance level of 5% for each transition, it is explainable that the most common combination only reaches a proportion of 0.85 for the log-rank-based test and 0.82 for the Cox method. This also explains why the HAJ estimators are performing nearly identical to the AJ estimator in this setting.

Another setting with similar results for the two tests is the partially frailty model. Both tests most frequently correctly detect only transition $2 \to 1$ as non-Markov. Thus, for this setting, more observations are kept overall,

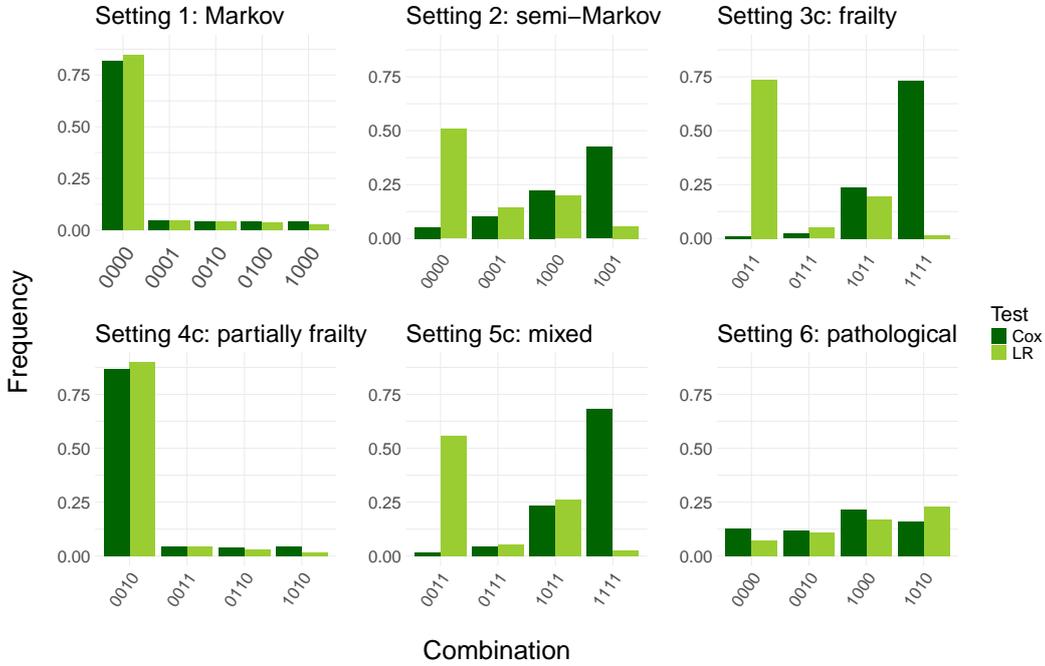

Figure C.11: Bar chart of the most common combinations of test decisions for all transitions for settings 1, 2, 3c, 4c, 5c and 6 for the log-rank-based test (LR) and the Cox method (Cox) over 10000 simulation runs for time point 245.57. Combinations are in the order $1 \to 2$, $1 \to 3$, $2 \to 1$ and $2 \to 3$, with 0 encoding a p-value $\geq 0.05$ and 1 a p-value $< 0.05$.

resulting in a favourable performance across the various measures while still applying landmarking where necessary.

For the other settings, the tests demonstrate different results. For both the frailty and the mixed model, the combination 0011 is most common for the log-rank-based test, meaning that it picks up non-Markov behaviour when transitioning out of state 2. The following most frequent combinations are 1011 and 0111. The correct combination 1111 is with 1.58% for setting 3c and 2.65% for setting 5c not common. In contrast, the Cox model predominantly identifies combination 1111, capturing all non-Markov transitions. This means, when looking at the estimator, the $\text{HAJ}^{\text{cox}}$ estimator will be equal to the LMAJ estimator in those cases. This result is reflected as nearly identical values for these two estimators in Figures 5–8 for setting 3c and 5c.

For setting 6, none of the tests has a clear most frequent combination.

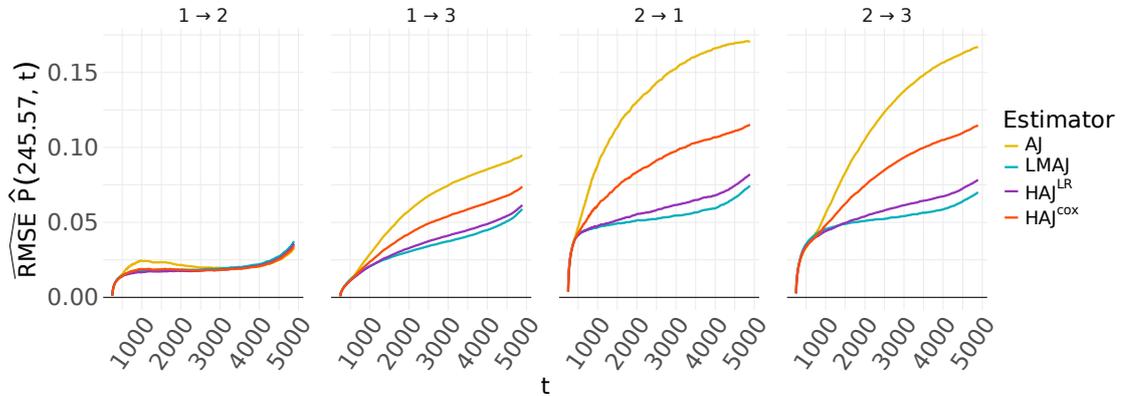

Figure C.12: RMSE for $\hat{P}_{hj}(245.57, t)$ for all transitions for setting 6: pathological, for the AJ, LMAJ, HAJ$^{\text{LR}}$ and HAJ$^{\text{cox}}$ estimator.

Both tend to find at most one non-Markov transition. Here, it needs to be considered, that the grid for the log-rank-based test did not include any time points before 40. These could have selected the right subjects to distinguish between the different parameters in the transition intensities and therefore, correctly reject the null hypothesis. Due to the interesting test results, this model is now investigated in more detail.

When examining the various measures for the pathological non-Markov model, the order of the estimators remains consistent across all transitions. The LMAJ estimator shows the best performance considering the RMSE (Figure C.12), bias (Figure C.13), and empirical coverage rate (Figure C.14), followed by the HAJ$^{\text{LR}}$, HAJ$^{\text{cox}}$, and AJ estimator in that order. An interesting observation for the empirical coverage rate is the unacceptably low values for both HAJ estimators, due to a high bias. In all other settings, the rates were usually much closer to the 0.95 level for these estimators. This is plausible, since the tests often only identified at most one transition as non-Markov. Considering all transitions, the LMAJ estimator emerges as the optimal choice for this setting. This example also underscores the importance of including all transitions, as the choice was clearer considering the newly added transitions compared to transition $1 \to 2$ evaluated in Section 5.

Consistent ordering of the estimators for all transitions across the different performance measures, as for the pathological process, is the case for most considered settings. But there are some exceptions with varying results

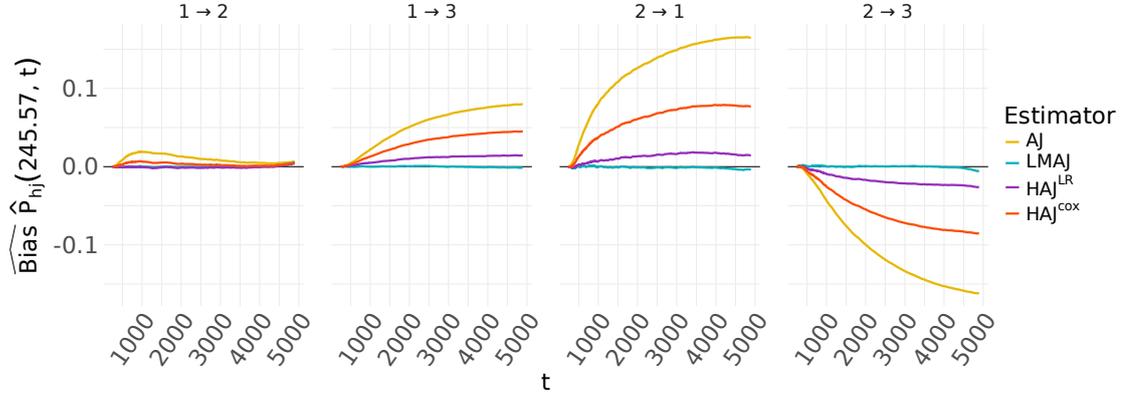

Figure C.13: Bias for $\hat{P}_{hj}(245.57, t)$ for all transitions setting 6: pathological, for the AJ, LMAJ, HAJ$^{\text{LR}}$ and HAJ$^{\text{cox}}$ estimator.

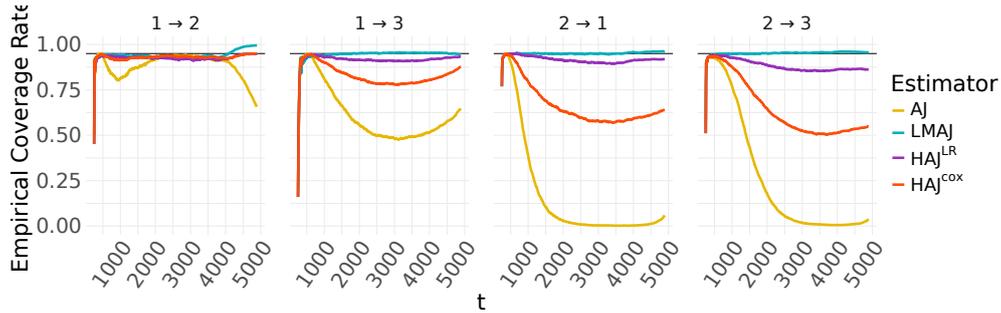

Figure C.14: Empirical coverage rate of the pointwise 95% confidence intervals for the transition probability $P_{hj}(245.57, t)$ for setting 6: pathological, for all transitions for the AJ, LMAJ, HAJ$^{\text{LR}}$ and HAJ$^{\text{cox}}$ estimator.

for different transitions. One of those exceptions is the previously mentioned setting 4b, depicted in Figure C.9. Here, there are different best-performing estimators depending on the transition, considering the RMSE. Another exception is the mixed process in setting 5c. The RMSE for all transitions is seen in Figure C.15. As shown in Section 5.3, for transition $1 \to 2$ the AJ estimator has a larger RMSE than the other estimators after the change from Markov to non-Markov simulation. This is, however, not true for the other transitions. The AJ estimator exhibits the lowest RMSE for all time points $t$ for transitions $1 \to 3$, $2 \to 1$, and $2 \to 3$. For transitions $1 \to 3$ and $2 \to 3$, this results from a smaller bias compared to transition $1 \to 2$ as depicted in

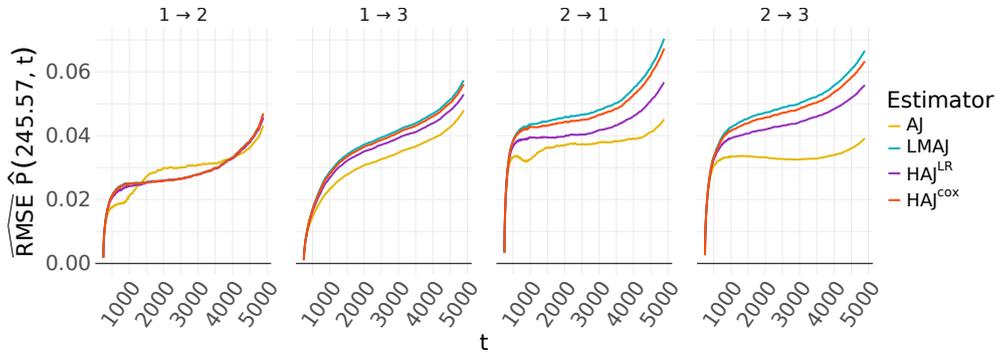

Figure C.15: RMSE for all transitions for $\hat{P}_{hj}(245.57, t)$ in setting 5c: mixed, for the AJ, LMAJ, HAJ$^{\text{LR}}$ and HAJ$^{\text{cox}}$ estimator.

Figure C.16. For transition $2 \to 1$, the AJ estimator seems to reduce the variance substantially.

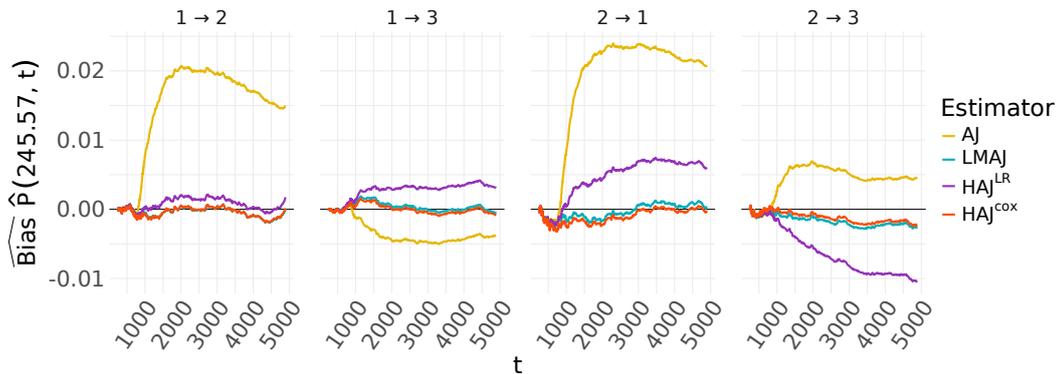

Figure C.16: Bias for all transitions for $\hat{P}_{hj}(245.57, t)$ in setting 5c: mixed, for the AJ, LMAJ, HAJ$^{\text{LR}}$ and HAJ$^{\text{cox}}$ estimator.

The differences between the transitions for this model are even more noticeable at the starting time 495.64. The corresponding RMSE is presented in Figure C.17. Here, in addition to transition $1 \to 2$, the AJ estimator also exhibits the highest RMSE for transition $2 \to 1$. This suggests, that this estimator may not be suitable for this scenario. Transitions $1 \to 3$ and $2 \to 3$, however, show opposite results with the lowest RMSE for the AJ estimator. For this model, therefore, no estimator is favoured considering the RMSE for all transitions. When the empirical coverage rate is included

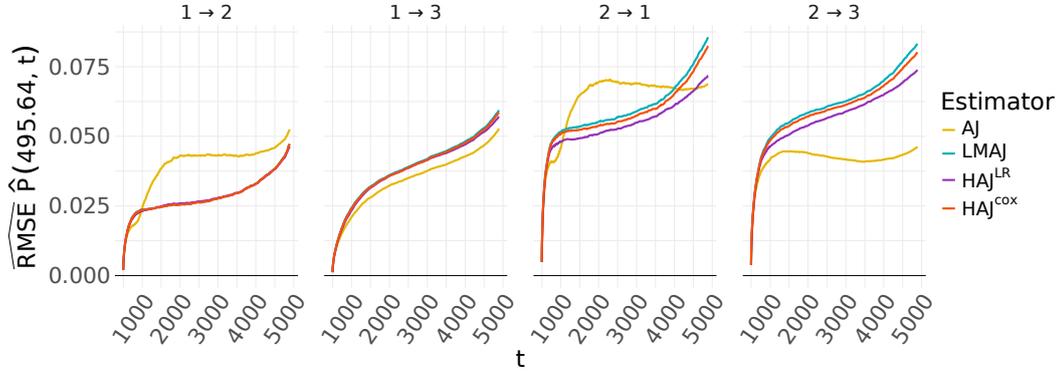

Figure C.17: RMSE for all transitions for $\hat{P}_{hj}(495.64, t)$ for setting 5c: mixed for the AJ, LMAJ, HAJ$^{\text{LR}}$ and HAJ$^{\text{cox}}$ estimator.

as a further criterion, depicted in in Figure C.18, the HAJ$^{\text{cox}}$ and LMAJ estimator yield higher values for all transitions close to 0.95, leading to a clearer decision favouring those two overall.

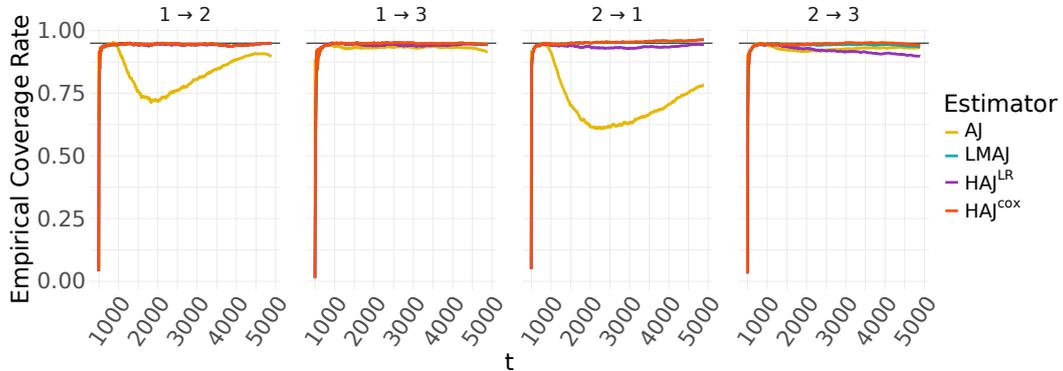

Figure C.18: Empirical coverage rate of the pointwise 95% confidence intervals for the transition probability $P_{hj}(495.64, t)$ for all transitions for setting 5c for the AJ, LMAJ, HAJ$^{\text{LR}}$ and HAJ$^{\text{cox}}$ estimator. The black horizontal line is at 0.95.

## C.4. Influence of the Starting Time

The following section explores various starting times $s$ and, therefore, landmark times, and their implications on the estimators. It is important to note that distinct starting times signify different transition probabilities with

varying interpretations. For instance, starting at earlier time points might capture events that occur more frequently, while starting at later time points might focus on events that occur later. Therefore, this section is intended solely for exploratory purposes rather than definitive conclusions. Its aim is to compare the differences between the estimators for the varying time points, rather than differences of the same estimator between those time points. To explore the influence of the starting time, three different time points, 0, 245.57, and 495.64, are considered.

Returning to the comparison of the empirical distribution of the p-values for the two test methods, depicted for the transition $1 \to 2$ in setting 3c in Figure C.19, reveals a slight decline in the detection of non-Markov transitions with increasing starting time point for both tests. This could probably be attributed to the decreasing number of subjects available for the analysis as the starting time point advances. Consistent with previous observations, the Cox method detects more non-Markov transitions across all time points than the log-rank-based test. Interestingly, when considering the performance measures for the estimators, this shift in test power is not apparent. The bias of the $\text{HAJ}^{\text{LR}}$ estimator actually decreases with increasing starting time point as seen in Figure C.20. This observation underscores again that the testing serves as a tool and is not an absolute determinant of the estimation outcomes.

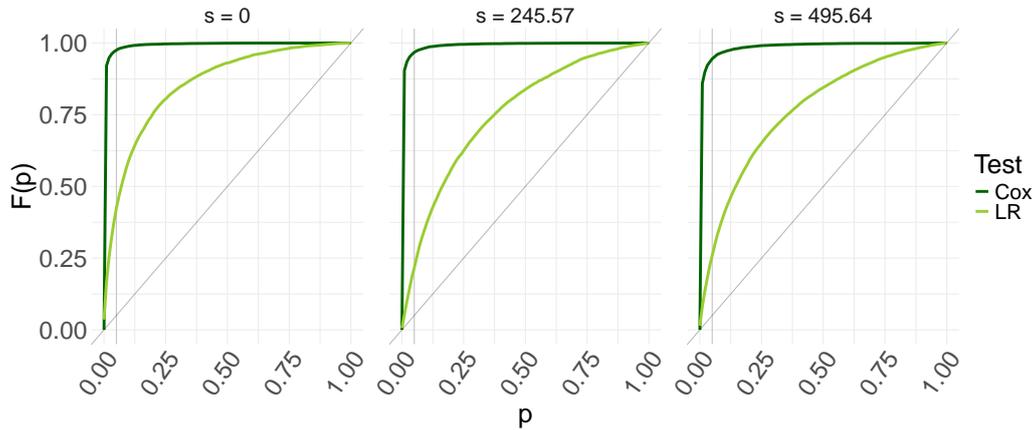

Figure C.19: Empirical distribution of the p-values for the log-rank-based Test (LR) and the Cox method (Cox) over 10000 simulation runs for setting 3c: frailty, with different starting time points for transition $1 \to 2$. The grey vertical line is at 0.05.

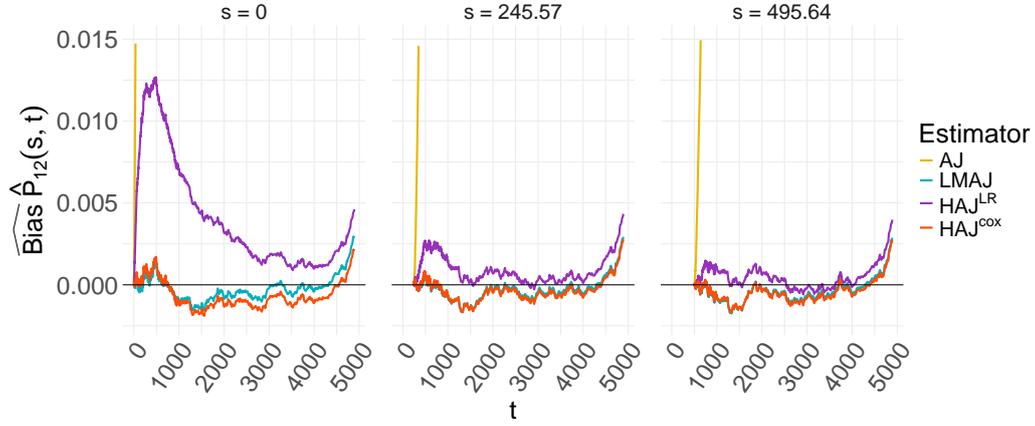

Figure C.20: Bias for $\hat{P}_{12}(s,t)$ for $s \in \{0, 245.57, 495.64\}$ in setting 3c: frailty, for the AJ, LMAJ, HAJ$^{\text{LR}}$ and HAJ$^{\text{cox}}$ estimator.

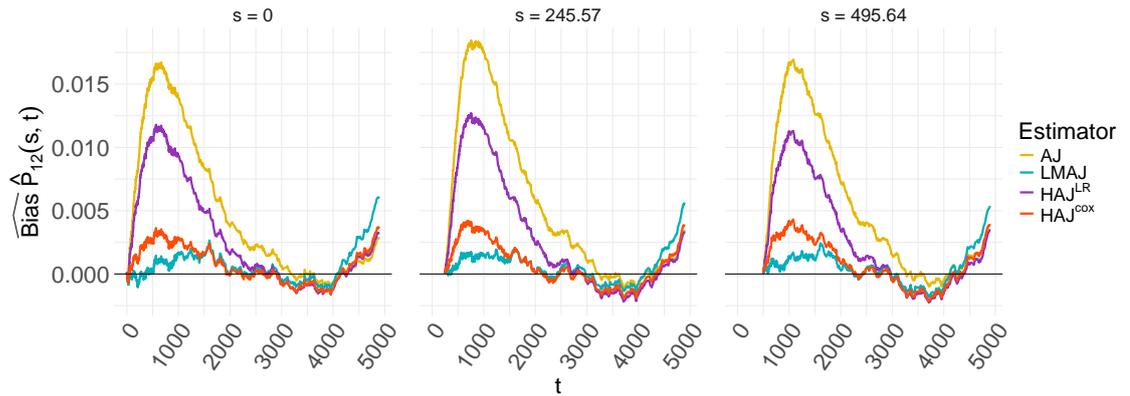

Figure C.21: Bias for $\hat{P}_{12}(s,t)$ for $s \in \{0, 245.57, 495.64\}$ in setting 2: semi-Markov, for the AJ, LMAJ, HAJ$^{\text{LR}}$ and HAJ$^{\text{cox}}$ estimator.

Upon examining the performance measures across the different settings, it becomes evident that, for the majority of settings, only slight differences are observed across various starting time points. Examples are depicted for the bias in setting 2 in Figure C.21 or for the RMSE in setting 6 in Figure C.22, where the values change, but the overall ranking and relative differences between the estimators tend to remain consistent. In some settings, a notable change is the overall increase in variance observed at later time points due to the reduced amount of available data. This is illustrated, for instance, for

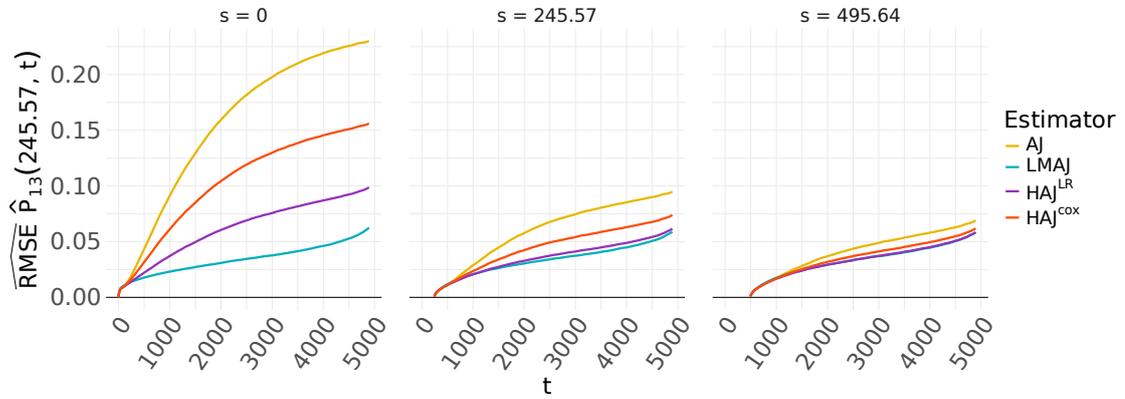

Figure C.22: RMSE for $\hat{P}_{13}(s,t)$ for $s \in \{0, 245.57, 495.64\}$ in setting 6: pathological, for the AJ, LMAJ, HAJ$^{\text{LR}}$ and HAJ$^{\text{cox}}$ estimator.

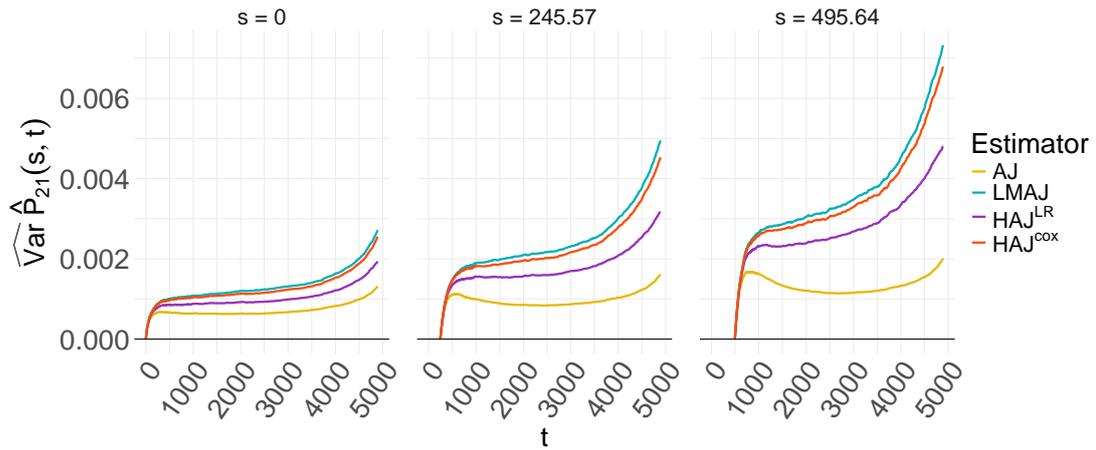

Figure C.23: Variance for $\hat{P}_{21}(s,t)$ for $s \in \{0, 245.57, 495.64\}$ in setting 5c: mixed, for the AJ, LMAJ, HAJ$^{\text{LR}}$ and HAJ$^{\text{cox}}$ estimator.

setting 5c in Figure C.23.

An example where the ranking of the estimators based on performance values changes, occurs in setting 5c, which was also the setting for the motivation of this section. The RMSE for transition $1 \to 2$ in this setting is depicted in Figure C.24. Initially, at starting time point $s = 0$, the AJ estimator yields the lowest values across all time points. However, with increasing starting time, the bias increases for the AJ estimator, consequently

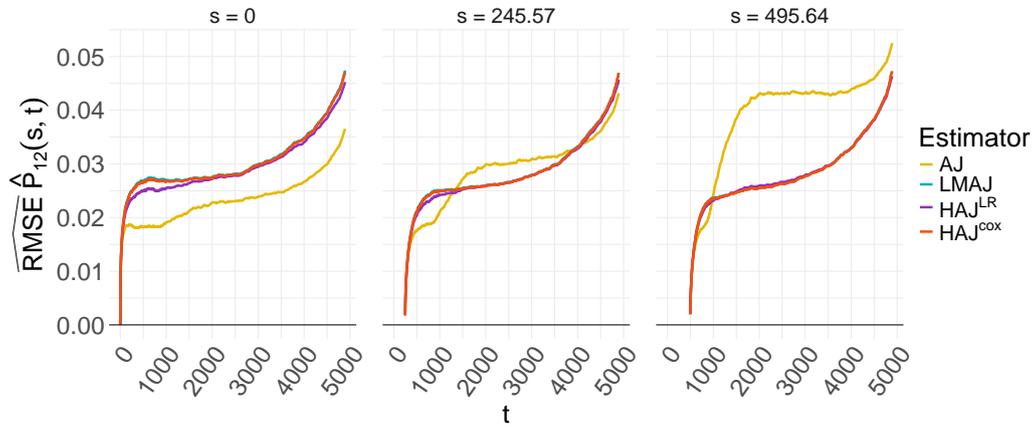

Figure C.24: RMSE for $\hat{P}_{12}(s,t)$ for different $s$ for setting 5c: mixed, for the AJ, LMAJ, HAJ$^{\text{LR}}$ and HAJ$^{\text{cox}}$ estimator.

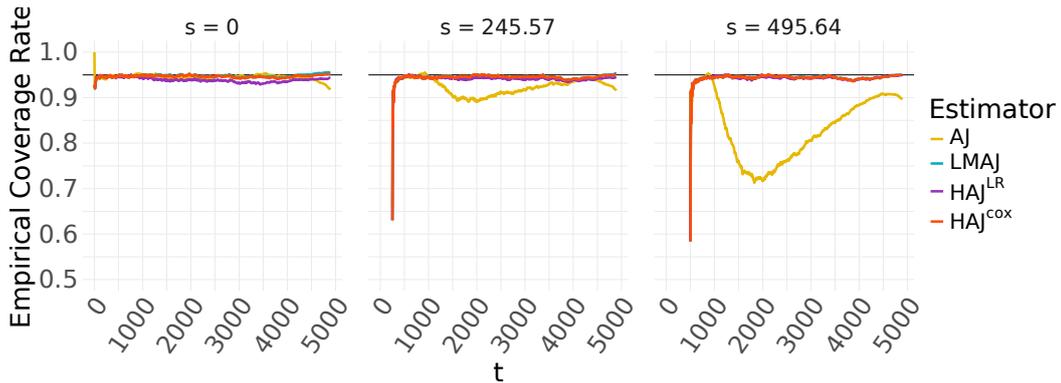

Figure C.25: Empirical coverage rate of the pointwise 95% confidence intervals for the transition probability $\hat{P}_{12}(s,t)$ for different $s$ for setting 5c: mixed for the AJ, LMAJ, HAJ$^{\text{LR}}$ and HAJ$^{\text{cox}}$ estimator.

leading to a higher RMSE than for the other estimators. This also explains why the empirical coverage rate decreases, depicted in Figure C.25. At $s = 0$, the rate is similar to the other estimators close to 95%. With increasing time points, the values drop down to less than 75%.

## D. Additional Figures

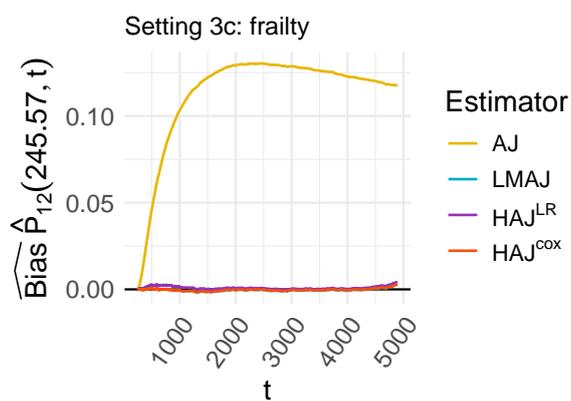

Figure D.1: Bias for $\hat{P}_{12}(245.57, t)$ for setting 3c for the AJ, LMAJ, HAJ$^{\text{LR}}$ and HAJ$^{\text{cox}}$ estimator.

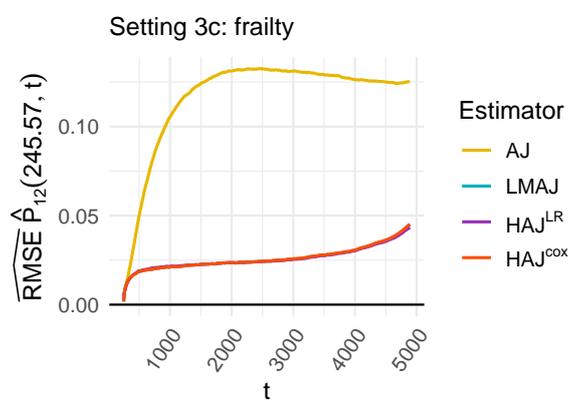

Figure D.2: RMSE for $P_{12}(245.57, t)$ for the setting 3c for the AJ, LMAJ, HAJ$^{\text{LR}}$ and HAJ$^{\text{cox}}$ estimator.

27

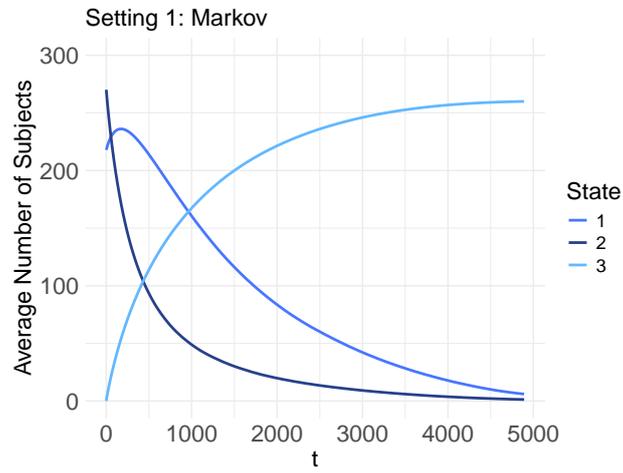

Figure D.3: Average number of subjects occupying each state for the different time points $t$ over the 10000 simulation runs for scenario 1: Markov.

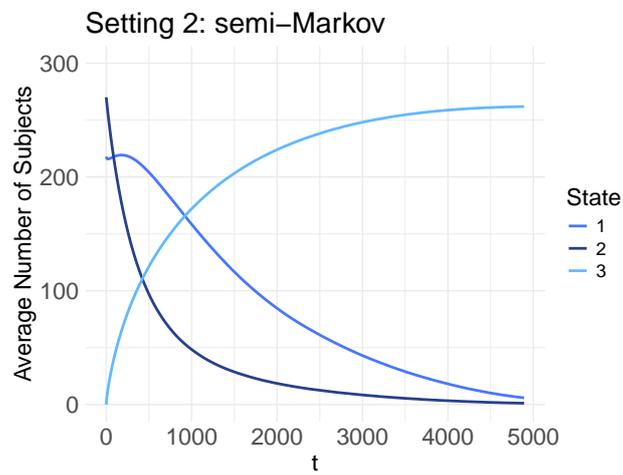

Figure D.4: Average number of subjects occupying each state for the different time points $t$ over the 10000 simulation runs for scenario 2: semi-Markov.

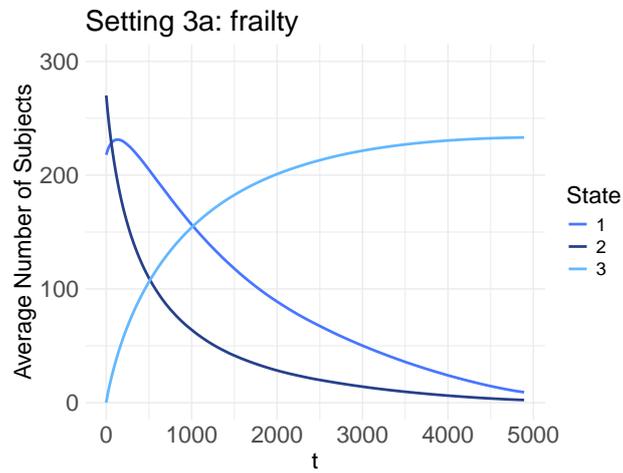

Figure D.5: Average number of subjects occupying each state for the different time points $t$ over the 10000 simulation runs for scenario 3a: frailty.

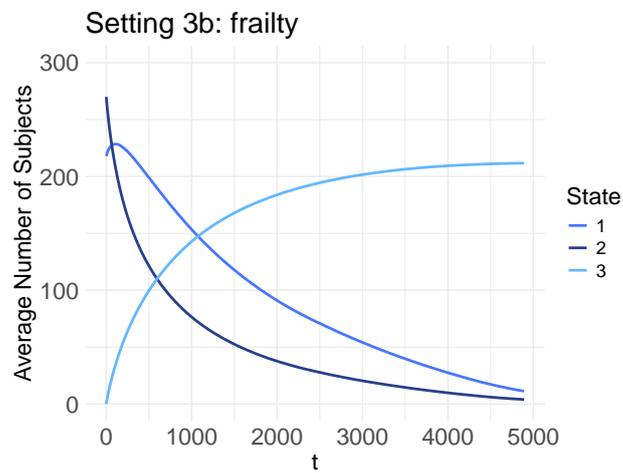

Figure D.6: Average number of subjects occupying each state for the different time points $t$ over the 10000 simulation runs for scenario 3b: frailty.

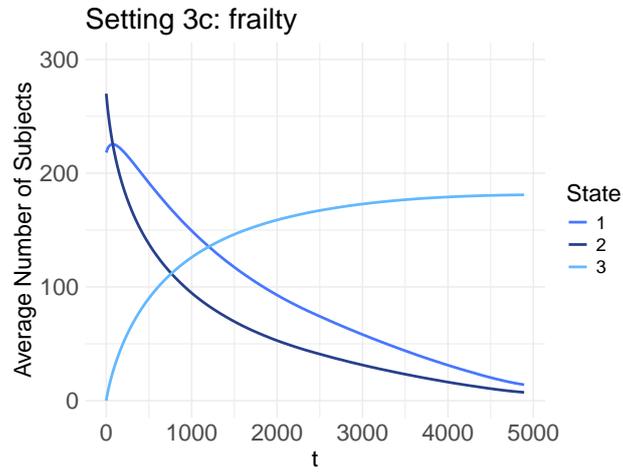

Figure D.7: Average number of subjects occupying each state for the different time points $t$ over the 10000 simulation runs for scenario 3c: frailty.

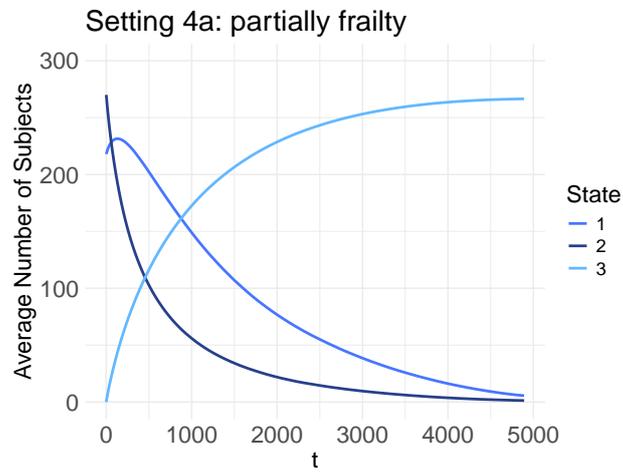

Figure D.8: Average number of subjects occupying each state for the different time points $t$ over the 10000 simulation runs for scenario 4a: partially frailty.

30

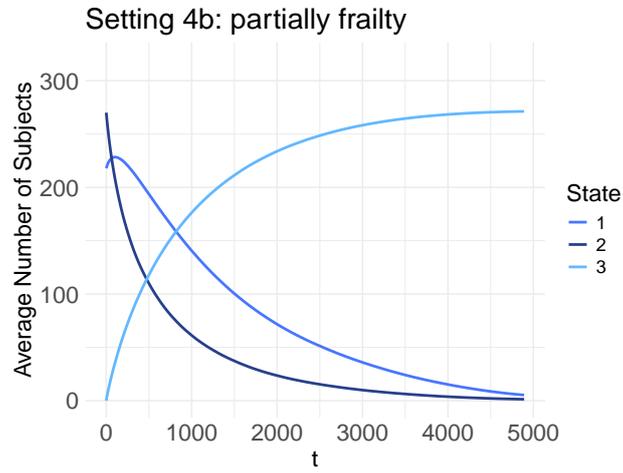

Figure D.9: Average number of subjects occupying each state for the different time points $t$ over the 10000 simulation runs for scenario 4b: partially frailty.

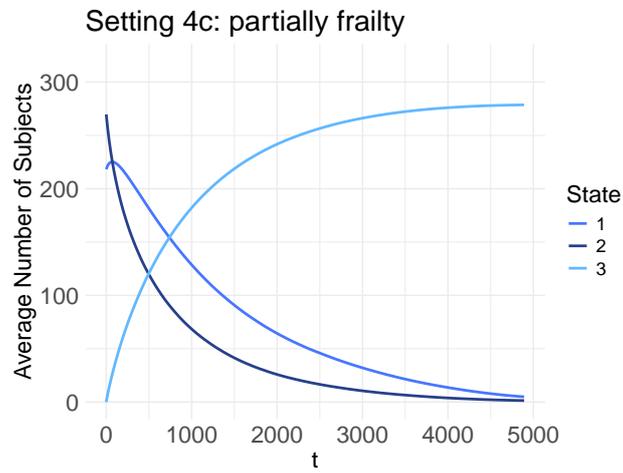

Figure D.10: Average number of subjects occupying each state for the different time points $t$ over the 10000 simulation runs for scenario 4c: partially frailty.

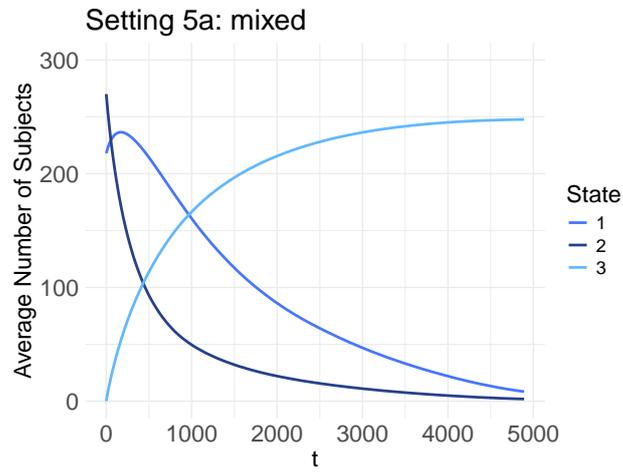

Figure D.11: Average number of subjects occupying each state for the different time points $t$ over the 10000 simulation runs for scenario 5a: mixed.

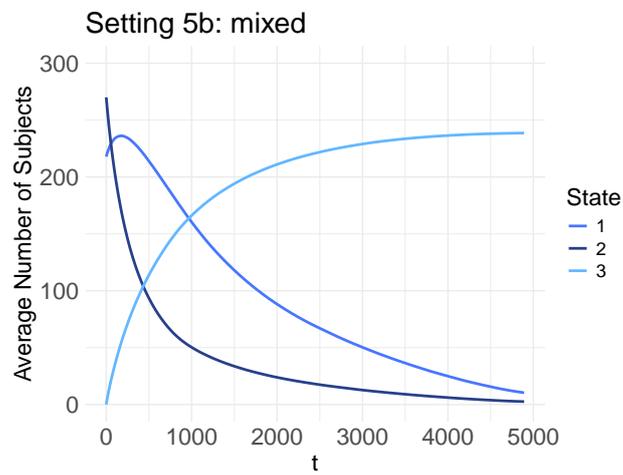

Figure D.12: Average number of subjects occupying each state for the different time points $t$ over the 10000 simulation runs for scenario 5b: mixed.

32

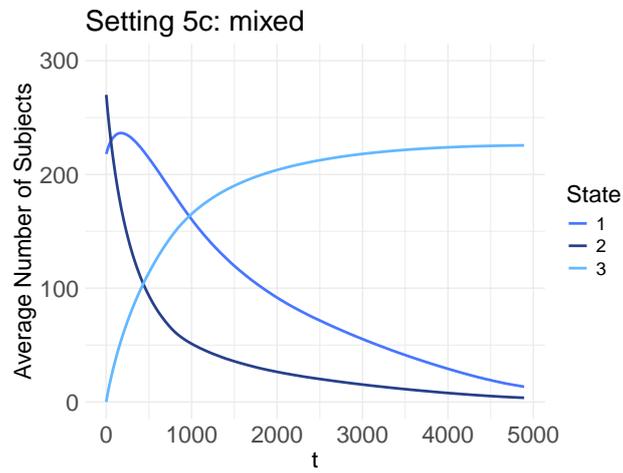

Figure D.13: Average number of subjects occupying each state for the different time points $t$ over the 10000 simulation runs for scenario 5c: mixed.

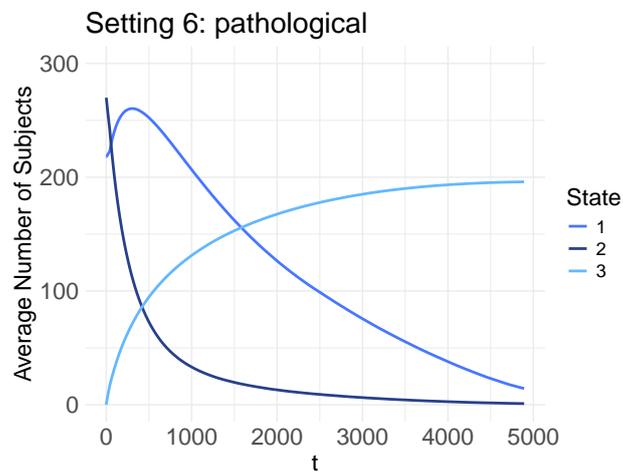

Figure D.14: Average number of subjects occupying each state for the different time points $t$ over the 10000 simulation runs for scenario 6: pathological.

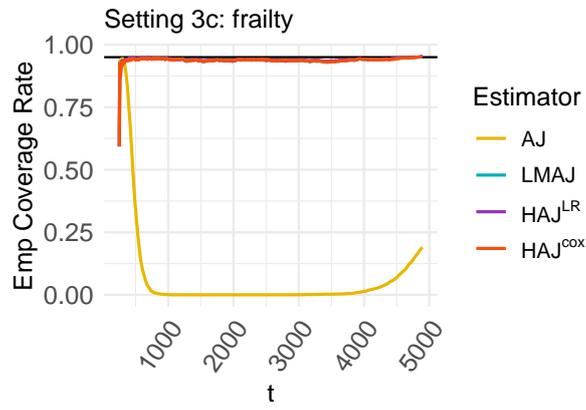

Figure D.15: Empirical coverage rate of the pointwise 95% confidence intervals for the transition probability $P_{12}(245.57, t)$ for the setting 3c for the AJ, LMAJ, $HAJ^{LR}$ and $HAJ^{cox}$ estimator. The black horizontal line is at 0.95.



\end{document}